\documentclass[twocolumn]{aastex631}

\usepackage{amsmath}
\usepackage[caption=false]{subfig}

\shorttitle{HSC shear catalog}
\shortauthors{Liu et al.}

\graphicspath{{./}{figures}}

\begin{document}

\title{Accurate Shear Recovery with Multi-Band Images of Hyper Suprime-Cam}

\author[0009-0006-2694-6752]{Cong Liu}
\affiliation{Department of Astronomy, Shanghai Jiao Tong University, Shanghai 200240, People's Republic of China}

\author{Jun Zhang}
\affiliation{Department of Astronomy, Shanghai Jiao Tong University, Shanghai 200240, People's Republic of China}
\affiliation{Shanghai Key Laboratory for Particle Physics and Cosmology, Shanghai 200240, People's Republic of China}
\email{betajzhang@sjtu.edu.cn}

\author{Hekun Li}
\affiliation{Shanghai Astronomical Observatory, Chinese Academy of Sciences, Shanghai 200030, People’s Republic of China}
\author{Pedro Alonso Vaquero}
\affiliation{Department of Astronomy, Shanghai Jiao Tong University, Shanghai 200240, People's Republic of China}
\author{Wenting Wang}
\affiliation{Department of Astronomy, Shanghai Jiao Tong University, Shanghai 200240, People's Republic of China}

\begin{abstract}
    The existing large scale weak lensing surveys typically reserve the best seeing conditions for a certain optical band to minimize shape measurement errors and maximize the number of usable background galaxies. This is because most popular shear measurement methods contain explicit or implicit thresholds on the galaxy-to-PSF (point spread function) size ratio, below which their shape measurement errors increase abruptly. Using the DECaLS data, we have previously demonstrated that the Fourier\_Quad method performs very well on poorly resolved galaxy images in general. It is therefore a ready tool for shear measurement with multi-band images regardless of their seeing conditions. In this paper, we apply the Fourier\_Quad pipeline on the multi-band images from the third public data release of the Hyper Suprime-Cam Subaru Strategic Program. We show that the shear catalogs from the five optical bands (g/r/i/z/y) all pass the field-distortion test with very high accuracy. Using the LOWZ and CMASS galaxies as foreground lenses, we show that the errorbar in the galaxy-galaxy lensing measurement can be decreased by factors around 15\% by combining shear catalogs from different bands. This indicates that it is worthful to do multi-bands shear measurements for a better shear statistics.
\end{abstract}


\section{Introduction}
Weak lensing refers to the coherent shape distortions of background galaxy images caused by the foreground large scale structure. Ever since its first detection \citep{2000MNRAS.318..625B,2000ApJ...537..555K}, it has become one of the major tools for probing the cosmic structure evolution, constraining the underlying cosmological models (especially $S_8$, \cite{DESY3cosmos,HSCY3cosmos,KiDS1000cosmos}), as well as exploring the properties of dark matter, dark energy, and neutrinos \citep{2008ARNPS..58...99H,2017MNRAS.467.4131V,LDP}.

The shape distortion due to weak lensing effect is tiny (typically 1\%), much smaller than the root mean square of the galaxy intrinsic ellipticity. Therefore, the weak lensing signals only show up in statistics, i.e., its accurate measurement requires a large sample of galaxy images. The ongoing Stage \uppercase\expandafter{\romannumeral3} weak lensing surveys include the Kilo-Degree Survey (KiDs\footnote{\url{https://kids.strw.leidenuniv.nl/}}, \cite{KiDssurvey}), the Dark Energy Survey (DES\footnote{\url{https://www.darkenergysurvey.org/}}, \cite{DESsurvey}), and the Hyper Suprime-Cam survey (HSC\footnote{\url{https://hsc-release.mtk.nao.ac.jp/}}, \cite{HSCsurvey}), each of which takes multi-band images of the sky that covers more than a thousand square degrees, providing a wealthy amount of high quality imaging data. 

Technically, the measurement of the weak lensing signal is challenging, as the galaxy morphologies are smeared by the point spread function (PSF), and pixelated on the CCD detector. They are also inevitably contaminated by the photon noise from the sky background and the detector. These effects are increasingly difficult to correct when the galaxy size becomes small comparing to the PSF size and the pixel size. For this reason, weak lensing surveys typically require good seeing conditions for galaxy shape measurement, so that more galaxies are well resolved and useful. The existing surveys therefore always reserve the best seeing condition for a certain band (or a few bands), based on which the galaxy shapes are measured, e.g. r-band for KiDs \citep{KiDs1000shear}, i-band for HSC \citep{HSCY3shear} and riz-bands for DES \citep{DESY3shear}.

Recently, the Fourier\_Quad (FQ hereafter) shear measurement method has demonstrated its ability in recovering acurrate shear information from faint/small sources with both simulations \citep{FQ_shear} and observational data \citep{FQ_fd,FQ_decals}. It is found that FQ can obtain reasonable shear catalog even from quite poorly resolved images (typically PSF FWHM $>$ 1 arcsec). We therefore expect FQ to be a promising tool to explore the additional shape information in multi-band imaging data. It was argued previously that additional images from multiple bands should not help much, as the galaxy shapes in different filters are highly correlated \citep{Jarvis}. But for the faint sources, as the photon noise becomes the main contributor to the shape measurement noise, we shall expect each additional exposure (regardless of the band) of the same galaxy to enhance the signal-to-noise ratio of the lensing signal to some extent. Moreover, independent multi-band shear measurement provides a natural consistency test of the overall quality of the shear catalog, and also a way of evaluating the imaging qualities in different optical bands, as shown in \cite{FQ_decals}.

In this work, we apply the FQ pipeline on the exposures of all available optical bands (g/r/i/z/y) from the third public data release of the Hyper Suprime-Cam Subaru Strategic Program (HSC). It has a high imaging depth (about 20 galaxies per square arcmin). Indeed, all five bands have reasonably good seeing conditions (0.61 arcsec on average for the best i band, and 0.79 arcsec for the worst g band). This makes HSC a very suitable sample for our multi-band shear measurement practice. In \S\ref{shear_measurement_method}, we give a brief introduction of the FQ shear measurement method. In \S\ref{image_processing}, we introduce the HSC data set, and describe the pipeline we use for the shape measurement of all the bands. We summarize the properties of our shear catalogs in \S\ref{shear_catalog_test}, and show the results from several consistency tests to quantify the multiplicative and additive bias. In \S\ref{performance}, we take galaxy-galaxy lensing as a tool to check the improvements when combining shear catalogs from different bands. Finally, we conclude in \S\ref{conclusion}. 

\section{Shear measurement method}
\label{shear_measurement_method}

In this work, galaxy shapes are estimated by the FQ method, a Fourier-space moment-based shear measurement algorithm. Its shear estimators are defined by the multipole moments of the galaxy power spectrum, which are written as:
\begin{equation}
    \begin{aligned}
    G_{1} &=-\frac{1}{2} \int d^{2} \vec{k}\left(k_{x}^{2}-k_{y}^{2}\right) T(\vec{k}) M(\vec{k}) \\
    G_{2} &=-\int d^{2} \vec{k} k_{x} k_{y} T(\vec{k}) M(\vec{k}) \\
    N &=\int d^{2} \vec{k}\left[k^{2}-\frac{\beta^{2}}{2} k^{4}\right] T(\vec{k}) M(\vec{k}) \\
    U &=- \frac{1}{2}\beta^{2} \int d^{2} \vec{k}\left(k_{x}^{4}-6 k_{x}^{2} k_{y}^{2}+k_{y}^{4}\right) T(\vec{k}) M(\vec{k}) \\
    V &=-2 \beta^{2} \int d^{2} \vec{k}\left(k_{x}^{3} k_{y}-k_{x} k_{y}^{3}\right) T(\vec{k}) M(\vec{k})
    \end{aligned}
    \label{eq:FQ_est}
\end{equation}
in which $\vec{k}$ is the wave vector. $M(\vec{k})$ is the galaxy power spectrum considering the correction of background noise and Poisson noise. $T(\vec{k})$ converts the galaxy PSF to a Gaussian form, i.e.:
\begin{equation}
    T(\vec{k})=\left|\widetilde{W}_{\beta}(\vec{k})\right|^{2} /\left|\widetilde{W}_{P S F}(\vec{k})\right|^{2}
\end{equation}
$\widetilde{W}_{P S F}(\vec{k})$ is the power of the PSF, and $\widetilde{W}_{\beta}(\vec{k})$ is the power of the isotropic Gaussian function $\widetilde{W}_{\beta}(\vec{x})=(2\pi\beta^2)^{-1}\exp(-|\vec{x}|^2/2\beta^2)$. The radius scale $\beta$ is selected to be slightly larger than the original PSF size to avoid singularities in the conversion. More details can be found in \cite{FQ_shear}.

It is demonstrated that the ensemble average of these estimators can achieve second order accuracy in shear recovery:
\begin{equation}
    \frac{\left\langle G_{1}\right\rangle}{\langle N\rangle}=g_{1}+O\left(g_{1,2}^{3}\right), \quad \frac{\left\langle G_{2}\right\rangle}{\langle N\rangle}=g_{2}+O\left(g_{1,2}^{3}\right)
    \label{eq:ave}
\end{equation}
Note that the averages are taken for $G_{1,2}$ and $N$ separately. An advantage of FQ is its good behavior for faint sources. 

The FQ shear estimators are proportional to the square of the galaxy flux, making it far from optimal to directly take their averages. As there are much more faint sources in survey, the distribution of shear estimators have a high peak and a long tail, leading to a large variance when taking assemble averages. \cite{FQ_pdf} proposed another way to measure shear statistics (called PDF-SYM method). The idea is to symmetrize the following probability distribution functions (PDF):
\begin{equation}
    \hat{G}_{1,2}=G_{1,2}- \hat{g}_{1,2}(N\pm U)
\end{equation}
When the distribution of $\hat{G}_{1,2}$ are best symmetrized with respect to zero, $\hat{g}_{1,2}$ becomes a good estimate of the true shear signal $g_{1,2}$. The quantity V is kept for the transformation of U under coordinate rotation. It is proved that the PDF-SYM method can automatically reach minimum statistical error (the Cramer-Rao Bound) in shear recovery.
It is also found that with some modifications, the PDF-SYM method can be applied in various different shear statistics, such as shear-shear correlation \citep{FQ_pdf}, galaxy-galaxy lensing \citep{Jiaqi_ggl,matt_ggl,pedro_ggl}, shear field reconstruction \citep{Haoran_pdf}. All the shear statistics in this work are calculated with the PSF-SYM method.

\section{Image processing pipeline}
\label{image_processing}

\subsection{HSC dataset}
\label{HSC_dataset}

Hyper Suprime-Cam (HSC) is a wide field optical camera on the 8.2m Subaru Telescope \citep{HSCsurvey}. Based on this camera, the Hyper Suprime-Cam Subaru Strategic Program (HSC) is a multi-layer survey containing Wide, Deep, and UltraDeep layers. The Wide layer aims to cover $1400 \deg^2$ of the sky with five ($g,r,i,z,y$) broad bands. The HSC survey has a unique advantage on the combination of imaging depth and resolution. The high image quality and multiple optical bands make it a powerful dataset for us to test the multi-band performance of shear measurement.

In this paper, we make use of the third public data release of HSC survey (\cite{hscdr3}, labelled as HSCpdr3 hereafter). HSCpdr3 contains 3810/3622/4625/4623/4346 exposures of images from g/r/i/z/y band, taken from Mar 2014 to Jan 2020. The sky coverage of each band is $\sim 1300 \deg^2$.
The images and data products of HSCpdr3 are produced by the HSC pipeline \citep{hscpipe}. They are all available for downloading from the HSC official website \footnote{\url{https://hsc-release.mtk.nao.ac.jp/}}. As our shape measurement pipeline requires single exposure images, we select the images with the prefix "CORR". This kind of images have accomplished several previous processes e.g. flatten field correction, background subtraction, defects detection.
In addition, we download another source catalog including RA, Dec, magnitudes of different bands, and photo-z. This catalog provides the photometric and redshift information for our final shear catalog. HSC team provides photo-z measured by different method \citep{HSCphotoz}. In this paper, we use the photo-z from the DEmP method \citep{demp}.

\subsection{Overview of the FQ Pipeline}
\label{FQ_pipeline}

Based on the FQ method, we developed a shear measurement pipeline which takes care of almost all the processes from the CCD images to shear catalogs. The pipeline starts with images after flat-field correction and bias correction. Our pipeline includes background subtraction, noise estimation, defects detection, and astrometric calibration. The sources are identified on individual chips and transformed into power spectrum. We then select star candidates and reconstruct PSF model all based on the power spectra of the source images. Finally, the shear estimators of galaxies are also calculated using the galaxy power sepctra in Fourier space. Among these steps, the astrometric calibration and star selection are carried out at the exposure level, and the other steps are processed at the individual chip level. 

FQ pipeline was previously applied on the Canada-France-Hawaii Telescope Lensing Survey (CFHTLens) dataset, the details of image processing are described in \cite{FQ_fd}. More recently, we update several parts of the pipeline and produce the shear catalog for the imaging data of the Dark Energy Camera Legacy Survey (DECaLS) \citep{FQ_decals}. The overall image processing for HSCpdr3 is quite similar to that of DECaLS. In the rest of this section, we only present the technical modifications that are special for HSCpdr3.

\subsection{Astrometric Calibration}
\label{Astrometric_Calibration}
The purpose of the astrometric calibration is to match the positions of the bright sources on CCD to those in the reference catalog (we use gaia \cite{gaiadr2}). The matched sources are used to recover the world coordinate system (WCS) parameters \citep{wcs2}. 

In the HSC ``CORR'' images, WCS system is recovered on the chip level. The reference point (CRVAL) is selected to be the center of each chip. But in our pipeline, we require a exposure level WCS system, in which all chips of same exposure share the same CRVALs. Thus we determine the CRVAL1,2 for the exposure by taking the averages of CRVAL1,2 of all the available chips. Note that in our case, this reference point is not forced to be the center of the exposure.    
The field distortion is quite significant on the HSC camera. If we only consider the basic WCS parameters, the pixel coordinates and the sky coordinates may suffer large offset especially near the edge of the focal plane, which makes it difficult in cross-matching the sources on the CCD with those in the reference catalog. Some chips may not find enough matched pairs for the fitting. Fortunately, HSC adopts the so-called Simple Imaging Polynomial (SIP) correction \citep{wcssip} to mitigate this issue. The correction is written as:  
\begin{equation}
    \left(\begin{array}{l}
    u \\
    v
    \end{array}\right)=\left(\begin{array}{ll}
    \mathrm{CD} 1_{-} 1 & \mathrm{CD} 1_{-}2 \\
    \mathrm{CD} 2_{-} 1 & \mathrm{CD} 2_{-}2
    \end{array}\right)\left(\begin{array}{c}
    x+f(x, y) \\
    y+g(x, y)
    \end{array}\right)
\end{equation}
Where $u,v$ is the intermediate world coordinate with the origin at (CRVAL1, CRVAL2). $x,y$ is the pixel coordinate with the origin at (CRPIX1, CRPIX2). CD i\_j is the matrix describing the linear transformation of the coordinates. $f(x,y)$ and $g(x,y)$ are the SIP corrections defined as:
\begin{equation}
    \begin{aligned}
    f(x, y)&=\sum_{p, q} \mathrm{~A}_{-} p_{-} q x^{p} y^{q}, \quad p+q \leq A\_ORDER \\
    g(x, y)&=\sum_{p, q} \mathrm{~B}_{-} p_{-} q x^{p} y^{q}, \quad p+q \leq B\_ORDER
    \end{aligned}
\end{equation}
Here $\mathrm{~A}_{-} p_{-} q$ and $\mathrm{~B}_{-} p_{-} q$ serve as the polynomial coefficients. The HSC pipeline applies A\_ORDER=B\_ORDER=3 for ``CORR'' images, and all these keywords can be found in the header of the fits file.
With these addition SIP correction, the stars from reference catalog are more accurately projected to the CCD plane.

\subsection{PSF reconstruction}

\begin{figure*}[htb]
    \centering
    \includegraphics[width=0.95\textwidth]{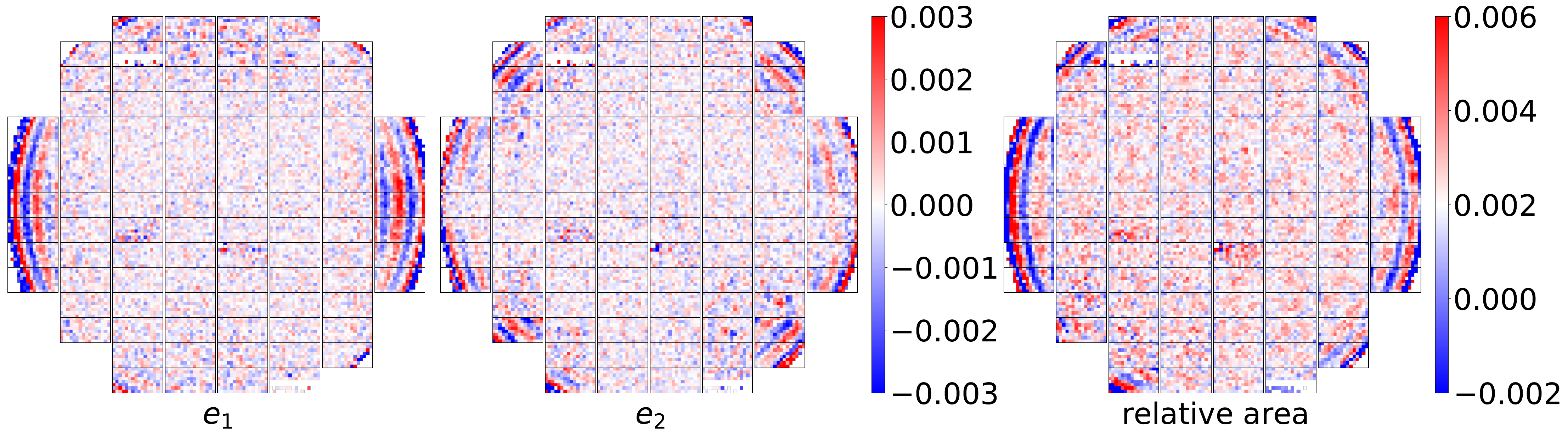}
    \caption{PSF residuals as a function of position on the focal plane. The panels from left to right are respectively $e_1, e_2$ and relative area residuals. These figures are calculated using the i-band data. Other bands have similar properties.}
    \label{fig:psf_res}
\end{figure*}

In the FQ pipeline, PSF reconstruction has two main steps. First, we pick up stars (and other point sources) from individual exposures as PSF models at their position. Then we fit a CCD level PSF field to get PSF at galaxy positions. For the star selection, our algorithm here basically follows \cite{FQ_fd}. The main steps are summarized below:

\begin{enumerate}
    \item First we collect bright sources (SNR $\ge$ 100) of the whole exposure as the star candidates. These candidates are then normalized by the total flux of the stamps, and then transformed into their power spectra.
    \item As the point sources have the most extended and similar profiles in Fourier space, the distribution of their sizes should form a major peak at the large end in the size distribution of the candidates. We locate the position of the peak, and estimate its Full Width at Half Maximum (FWHM)(or $\sigma$). We exclude the candidates that are more than 3-$\sigma$ away from the peak.
    \item Then we build a reference PSF model from the remaining candidates. For each pixel of the PSF model, we sort the corresponding pixel value of all the star candidates and use the value ranked at 25\% from the top. We study the similarity between the star candidates and the PSF model. The similarity is defined as the $\chi^2$ of the two stamps:
    \begin{equation}
        \chi^2 = \Sigma_i (I_1^i-I_2^i)^2/(I_1^i+I_2^i)
    \end{equation}
    Here $I^i$ refers to the $i^{th}$ pixel of the image (power spectrum). The distribution of $\chi^2$ also contains a peak at the small end. We again locate the position of the peak, and estimate its FWHM/$\sigma$. The candidates that are more than 3-$\sigma$ away from the peak are removed.
    \item Finally, we build an exposure-wide PSF model by fitting a polynomial function up to the $9^{th}$ order using all the remaining star candidates. This time we calculate the $\chi^2$ between the remaining candidates and the PSF models at their positions. We again remove outliers whose $\chi^2$ are more than 3-$\sigma$ away from the peak of the $\chi^2$ distribution. All the survived candidates are treated as the true point sources.
\end{enumerate}

Our final PSF model is constructed by fitting polynomial functions on the chip level. Note that in the FQ method, we directly recover the power spectrum of PSF instead of its image in real space. This is indeed an advantage because power spectrum is automatically centered. The reconstructed PSF power spectrum $W_{PSF}(u,v)$ at any CCD coordinate x,y is written as:
\begin{equation}
    \left. W_{PSF}(u,v)\right\vert_{x,y}=\sum_{i=0}^n\sum_{j=0}^i A_{i,j}(u,v)x^{j}y^{i-j}
\end{equation}
Here "n" refers to highest polynomial order. We choose $n=3$ for the best performance. The fitting is carried out pixel-by-pixel ($48\times48$).

In fig.\ref{fig:psf_res}, we show the average PSF residuals of ellipticity ($e_1,e_2$) and relative area on the focal plane. The results are carried out by comparing the morphology of stars and recovered PSFs at the position of stars. Here, $e_1,e_2$ and area are defined in Fourier space as:
\begin{equation}
    \begin{aligned}
        e_1 = \frac{Q_{20}-Q_{02}}{Q_{20}+Q_{02}} \\ 
        e_2 = \frac{2 Q_{11}}{Q_{20}+Q_{02}} \\ 
        area = \frac{Q_{20}+Q_{02}}{Q_{00}} 
    \end{aligned}
\end{equation}
Here, $Q_{ij}=\int k_x^ik_y^jP(\vec{k})d^2\vec{k}$ are the moments in the Fourier space. $P(\vec{k})$ is the power spectrum of the PSF. In order to reject the contribution of high frequency noise, we only use region with $P(\vec{k})>0.02P(0)$ in the integrals. 

The residual patterns are quite similar to the PSF reconstruction results by the official HSC pipeline (see fig.9 of \cite{hscpipe}). There are obvious ring-like feature at the outer regions of the exposure. This feature appears in all five bands. Increasing order of the polynomial function can somewhat suppress these features. But considering the limited star numbers on CCD, we still fix $n=3$ to avoid overfitting problem. As we stack all exposures together, this ring-like residuals tend to be related with the PSF caused by optical instruments. This fact suggests that we could study this residual structure as a function of position on focal plane using all star candidates. Such an improved PSF reconstruction method is discussed in our another work \citep{alonso2024}.   

\subsection{Other Issues}
\label{other_issue}

In the FQ pipeline, the defects detection part includes several subroutines to locate different types of defects, e.g. bad pixels, cosmic rays, stripes, halos... The HSC ``CORR'' images have already detected and decorated defects. We make use of the official mask files. Meanwhile, we also carry out our own defects detection in order to mask unfound defects, such as some stripe-like features along the read-out direction of the CCD, which are likely caused by problems related to charge transfer. The final catalog of the defects includes those from both the official mask files and our own algorithm.

Regarding the deblending issue, we adopt a simple algorithm in our pipeline: We throw away spatially connected galaxy pairs if their redshift difference is larger than 0.1. Otherwise we treat them as a single source. After applying this algorithm, we remove about 6.58\% of the galaxies in our final shear catalogs.  

\begin{figure}[htb!]
    \centering
    \includegraphics[width=\columnwidth]{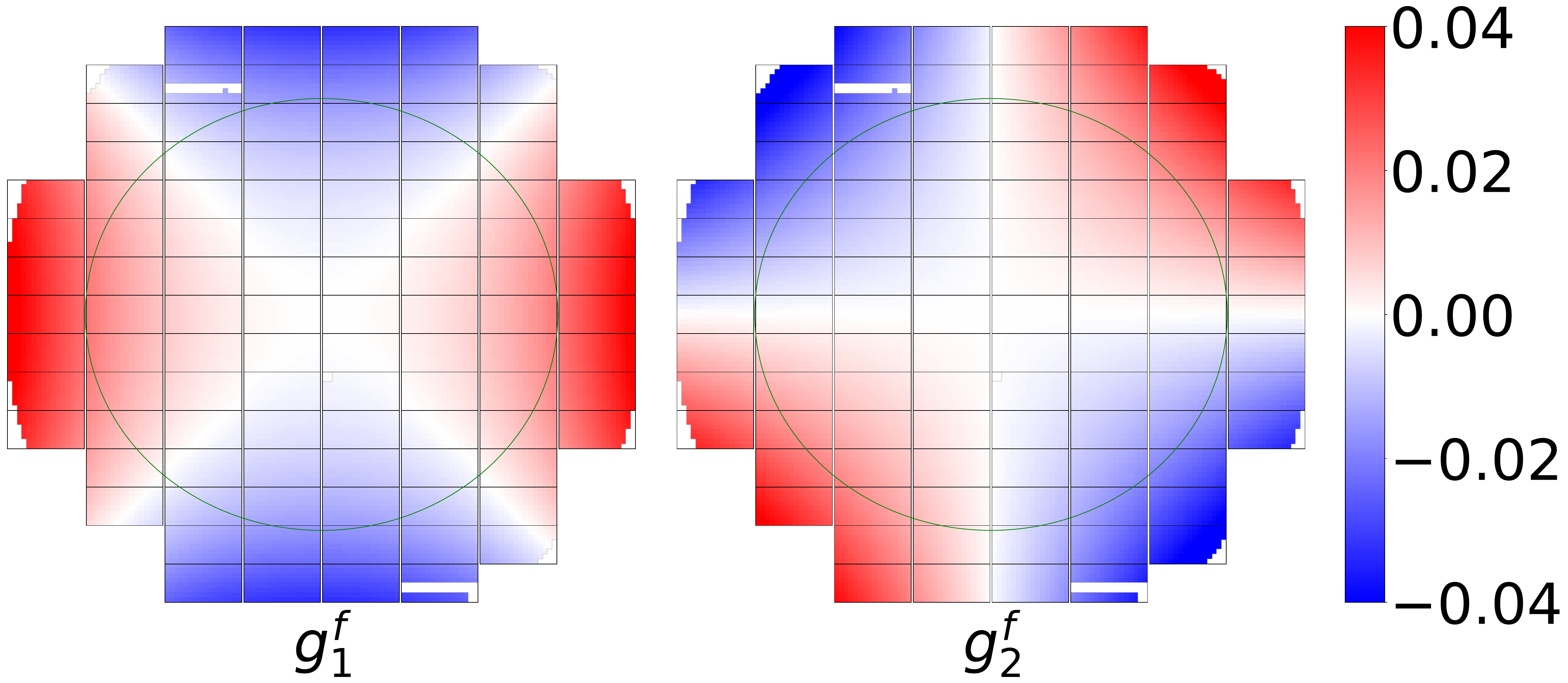}
    \includegraphics[width=0.6\columnwidth]{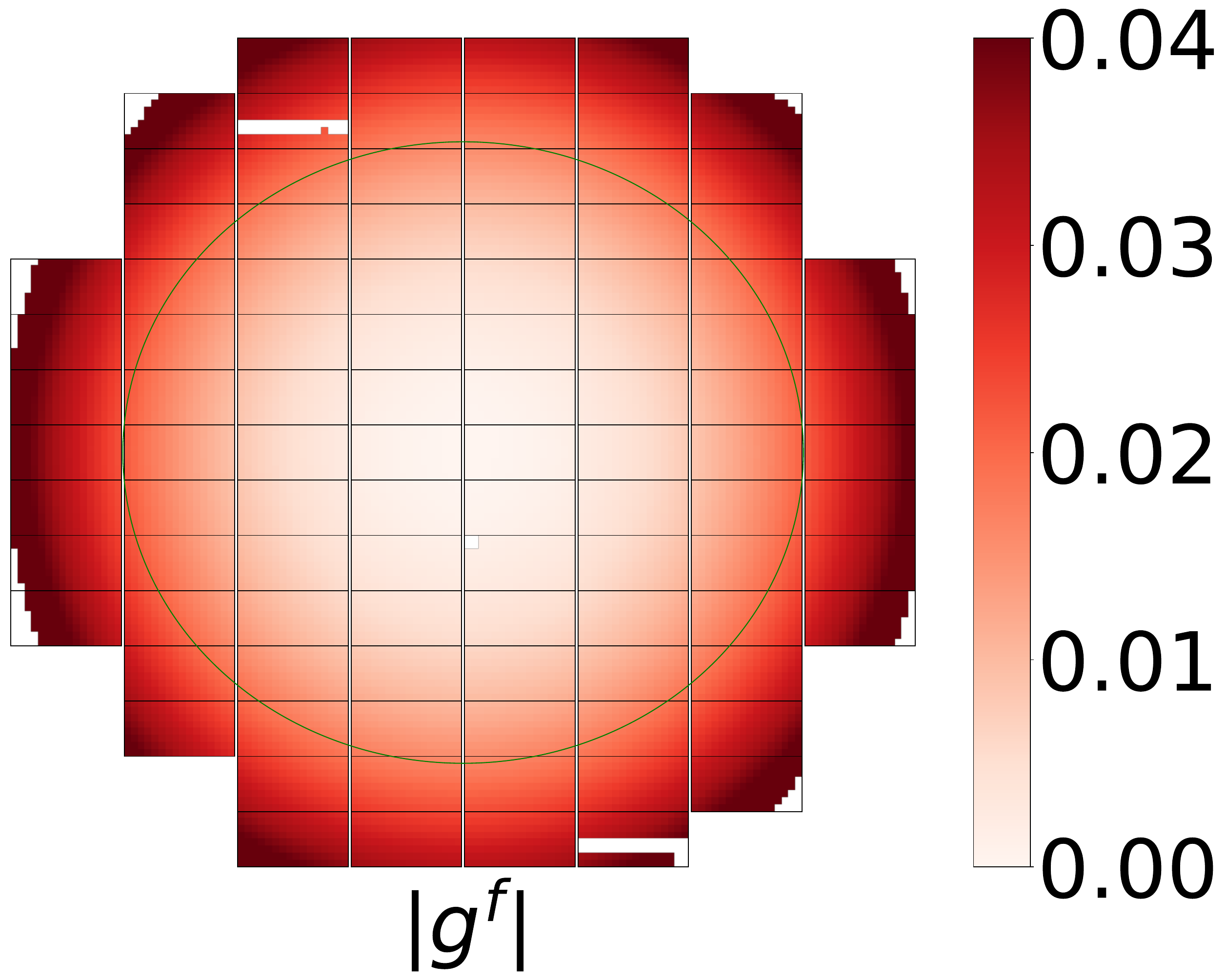}
    \caption{The field distortion shear distribution recovered by astrometric calibration. Black boxes represent the CCD arrangement. The upper panels show the distributions of the two shear components of FD: $g^f_1$ and $g^f_2$. The lower panel shows the total shear amplitude, $|g^f|=\sqrt{(g^f_1)^2+(g^f_2)^2}$. Green lines in each panel represent the position where $g^f=0.02$. This circle defines our preferred boundary of high-quality shear catalog.}
    \label{fig:gf_dist}
\end{figure}

\section{Multi-Band Shear Catalog}
\label{shear_catalog_test}

In this section, we check the performance of our shear catalogs by carrying out two tests: the field distortion test and the focal plane null test. We show how to make cuts in our shear catalogs to avoid systematic biases. We also briefly summarize several properties of our catalogs in the end.

\subsection{Field Distortion Test}

The field distortion effect causes the shape distortion of the observed object, introducing an equivalent shear (label as field distortion shear, $g^f_1,g^f_2$, hereafter). \cite{FQ_fd} shows that the field distortion shear provides a convenient way to test the accuracy of shear measurement and estimate the multiplicative and additive biases from the observational data directly.

The field distortion effect can be described by a series of WCS parameters, which are recovered during astrometric calibration. Fig.\ref{fig:gf_dist} shows the recovered field distortion shear of a typical HSC exposure. As indicated in this figure, the field distortion shear signal is a function of position on the focal plane, and its value is comparable to the lensing shear. 

\begin{figure*}[htb!]
    \centering
    \includegraphics[width=0.95\textwidth]{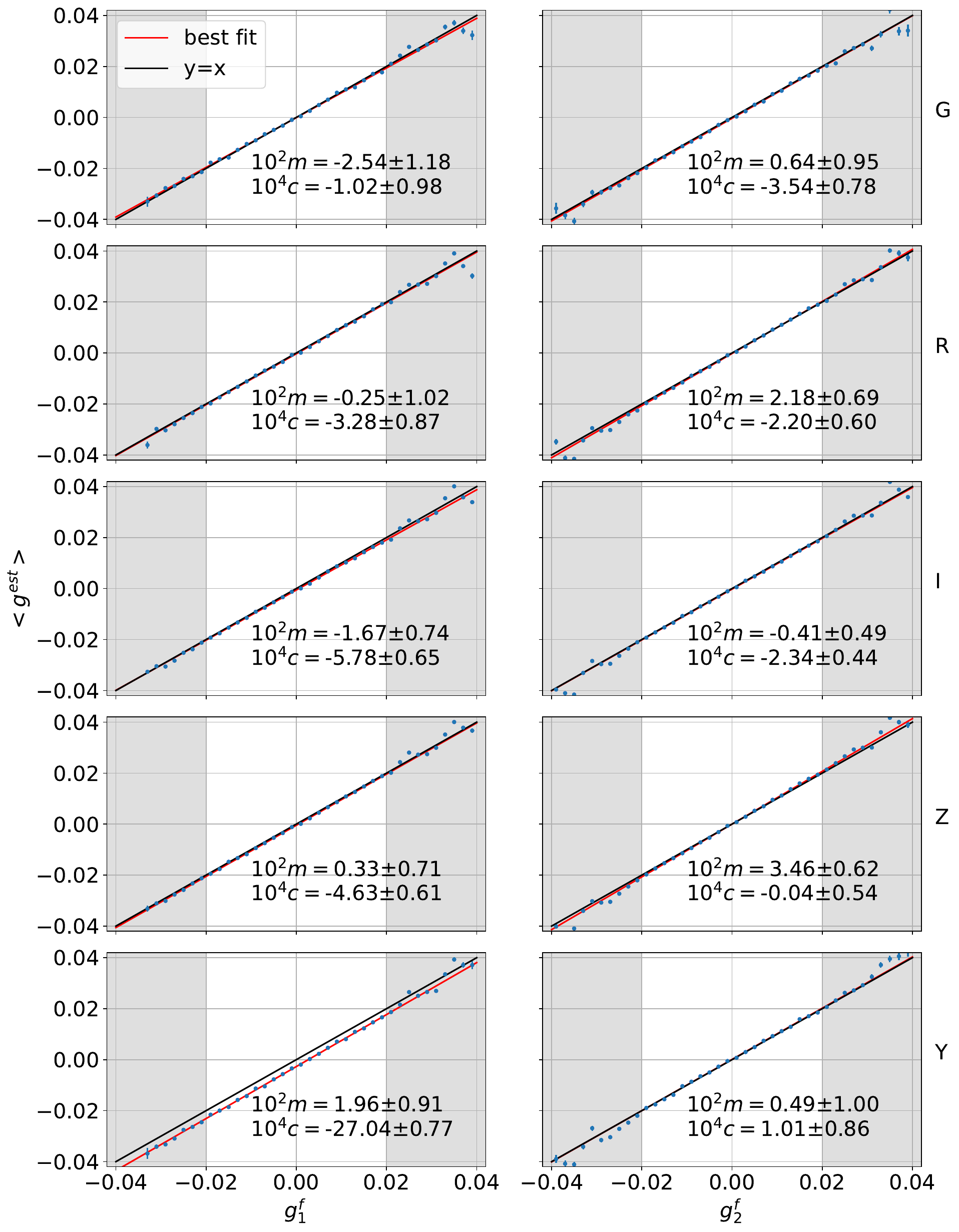}
    \caption{The results of the field distortion tests for different bands (g/r/i/z/y) and shear components ($g_1$,$g_2$). In each panel, the blue data points (with small error bars) represent the stacked shear signals from galaxies binned according to their field distortion shear values ($g^f_1$ and $g^f_2$). The red lines show the best-fit results. The values of the multiplicative and additive biases are given at the bottom of each panel.}
    \label{fig:fd_mc}
\end{figure*}

If we stack the measured galaxy shear estimators (without removing the field distortion shear) with same field distortion shears, the result is supposed to be consistent with $g^f$ itself. And considering the recovered shear as a function of the field distortion shear, it is easy to estimate the linear measurement bias by:
\begin{equation}
    g_{1,2}^{est}=(1+m)g_{1,2}^f + c
\end{equation}
The main results of the test are shown in fig.\ref{fig:fd_mc}. Note that each blue data point in the figure does contain an error bar that is very small. As shown in the plot, there are wiggles at the regions with large field distortion shear (i.e. the outskirts of the focal plane). This is apparently caused by the PSF residuals with a ring-like pattern shown in fig.\ref{fig:psf_res}. It is a direct evidence of how strongly the PSF error correlates with the shear error. As the wiggles almost vanish at the inner parts ($|g^f_{1,2}|\le0.02$). Thus we decide to make a cut with $\vert g^f\vert\le0.02$ in our shear catalogs\footnote{$\vert g^f\vert=\sqrt{(g_1^f)^2+(g_2^f)^2}$}. The remaining area on the focal plane is inside the green circle shown in fig.\ref{fig:gf_dist}. 

In addition, it is also necessary to consider a cut at the faint end of the galaxy using, e.g., SNR, apparent magnitude, resolution factor. \cite{snrf} shows that these selection functions might introduce selection bias for faint sources. A low bias selection factor is proposed in this work, which is conveniently defined in Fourier space:
\begin{equation}
\nu_F = \frac{P(\vec{k}=0)}{\sqrt{N}\sigma}
\end{equation} 
Here $P(\vec{k}=0)$ is the central point of image power spectrum, i.e. the total flux of stamp. N is the stamp size, and $\sigma$ is the rms of background noise. In this work, we adopt this factor to make cut on the galaxy sample. We find that $\nu_F>5$ is a good choice considering the bias level and the galaxy number. 

As shown in these plots, all five bands have percent level multiplicative bias, and these m are consistent with zero within $~3\sigma$ level. For the additive bias, there is an obvious c appearing in the y band $g_1$ panel. This might has something to do with the y band scatter light problem (see \cite{hscdr2} for detail). We do not focus on it too much. An additional constant shear is added in the final shear catalog to correct for it. Other additive biases are all found to be at the level of a few times $10^{-4}$ or less. It is worth noting that these results are achieved without any calibrations from image simulations. It demonstrates the ability of our FQ pipeline in obtaining precise multi-band shear catalog from the HSCpdr3 dataset.

\subsection{Focal Plane Null Test}
\label{null_test}

Similar to the study of the PSF residuals, we measure the average shear signals/residuals as a function of the location on the focal plane. The null test allows us to locate the problematic regions on the CCDs. For this purpose, each CCD is divided into 8*16 grid cells, and we stack the shear signals in each cell.
\begin{figure*}[htb!]
    \centering
    \includegraphics[width=0.95\textwidth]{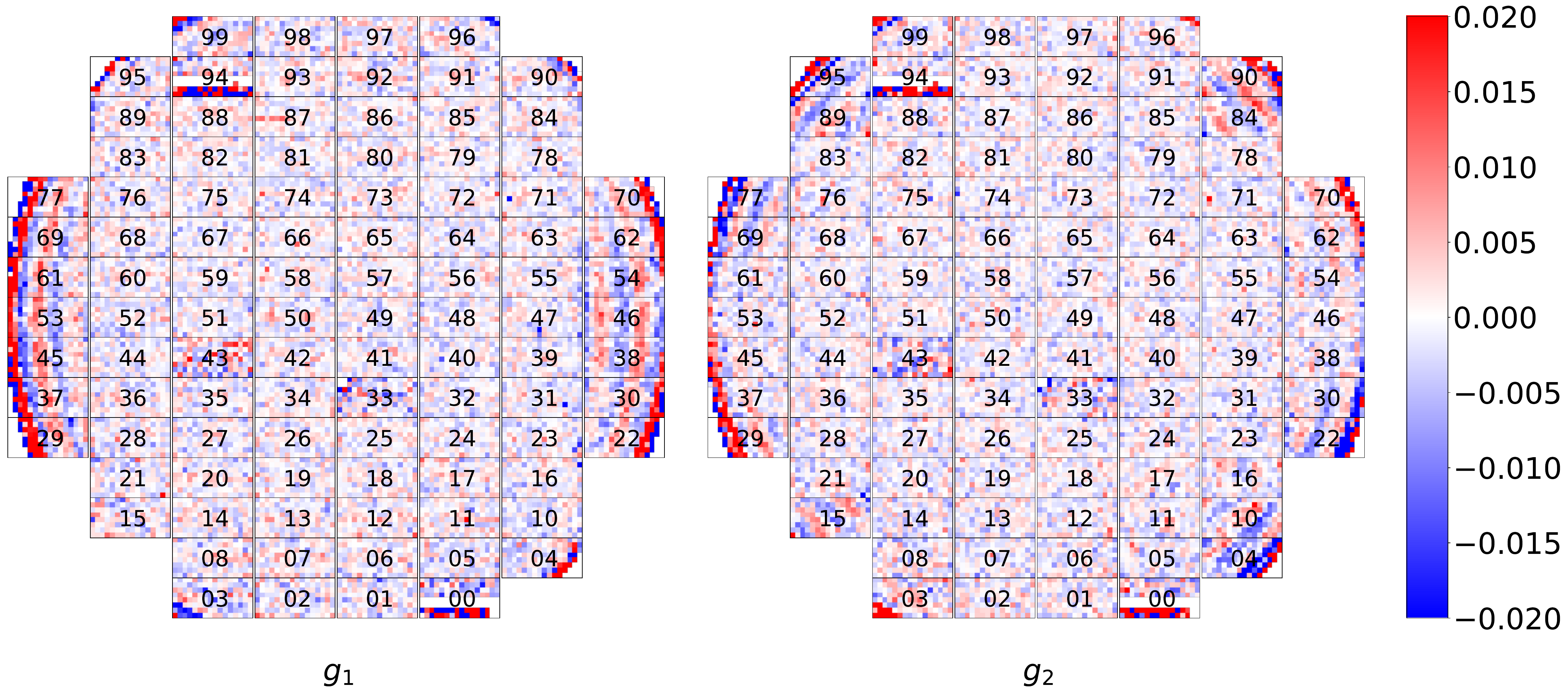}
    \caption{Results of the focal plane null test for the i-band shear catalog. The left and right panels are for $g_1$ and $g_2$ respectively. Results of the other bands have similar properties.}
    \label{fig:2d_fd_test}
\end{figure*}
Fig.\ref{fig:2d_fd_test} shows the results of the test. It only presents the i-band result. The other bands show similar patterns. We can again see the ring-like features towards the edges of the exposure, which are highly correlated with the PSF residual patterns shown in fig.\ref{fig:psf_res}. Besides this global feature, it is clear to see that the CCDs with chip id 0, 33, 43, 94 are obviously too noisy, even containing some blank regions. This feature is related with the bad channels on these CCDs, which are reported by the HSC team \footnote{\url{https://hsc.mtk.nao.ac.jp/pipedoc/pipedoc\_8\_e/hsc\_info\_e/index.html\#hsc-info}}. We show some examples from these four CCDs in fig.\ref{fig:badCCD}. Note that although the problematic areas in these chips are masked as defects, large amount of masked pixels could still affect fittings on the chip scale, such as background estimation and PSF modelling, resulting in shear measurement bias. Thus we decide to remove these four CCDs from our shear catalogs. Note that in making fig.\ref{fig:fd_mc}, the shear catalog from these CCDs are not included.

\subsection{PSF polynomial order}

As shown in fig.\ref{fig:psf_res}, fig.\ref{fig:fd_mc}, and fig.\ref{fig:2d_fd_test}, the PSF modelling directly influence the accuracy of shear measurement. One can therefore directly rely on the field distortion test to study the performance of PSF reconstruction. Here we briefly discuss the choice of the polynomial order in PSF modelling. 

We adopt the polynomial functions of order $n=1,2,3$ to model the spatial variation of the PSF power spectrum. The results are shown in fig.\ref{fig:wiggle}, in which the field distortion shear has been subtracted from the stacked shear to enhance the contrast between different cases. One can see in the figure that the residuals/wiggles on the edge of the focal plane ($\vert g_i\vert >0.02$) are significantly suppressed when we increase n. Although one could further increase n beyond 3 to suppress the wiggles, it quickly becomes impractical given the limited number of bright stars on each CCD. Instead, we simply choose $n=3$ in this work, and remove the shear catalog in the region of $\vert g^f\vert >0.02$.

\subsection{Shear Catalog Properties}
\begin{figure}[htb!]
    \centering
    \includegraphics[width=\columnwidth]{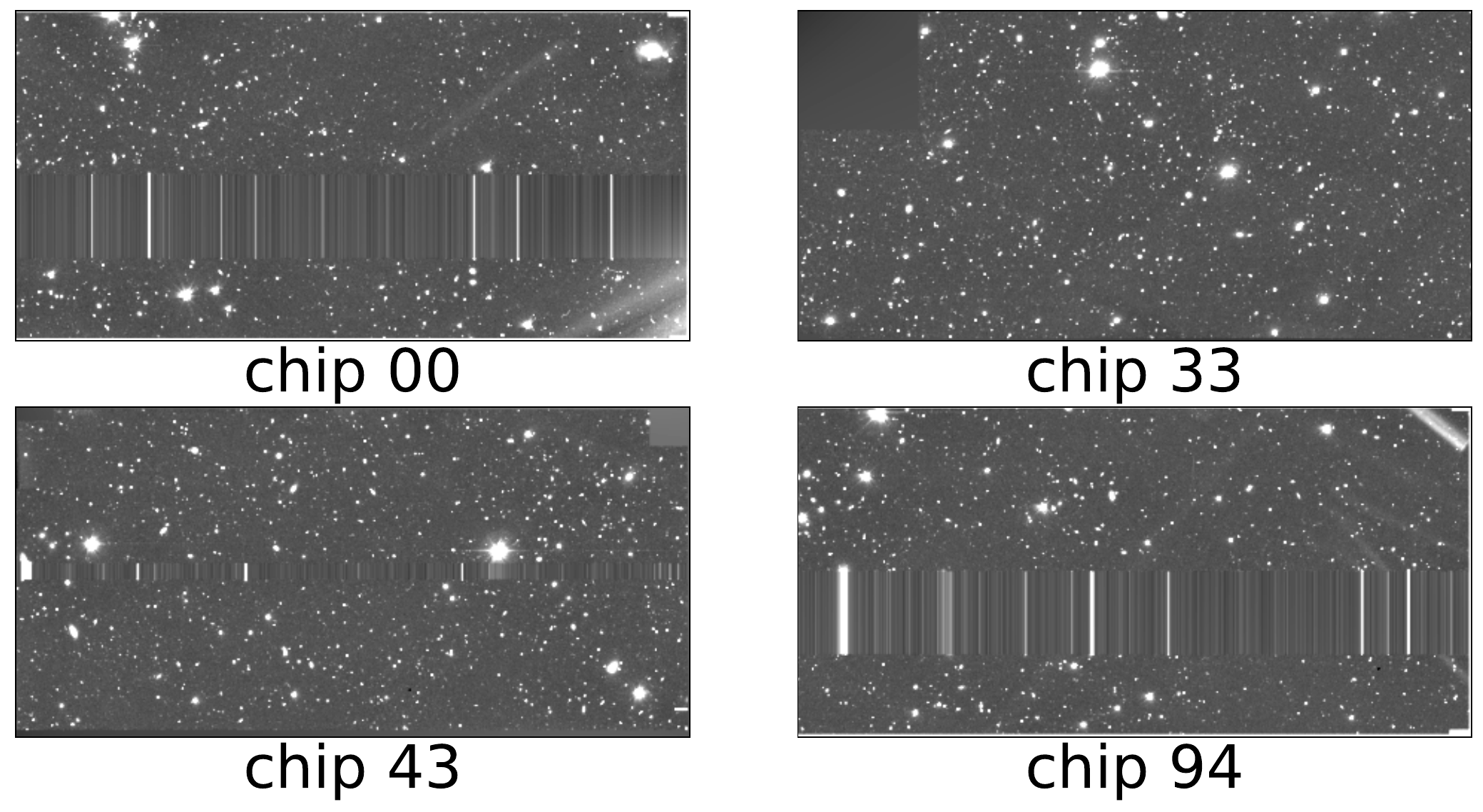}
    \caption{Examples of bad channels on CCD. This figure shows the images with chip id 0, 33, 43, 94. }
    \label{fig:badCCD}
\end{figure}
\begin{figure}[htb!]
    \centering
    \includegraphics[width=\columnwidth]{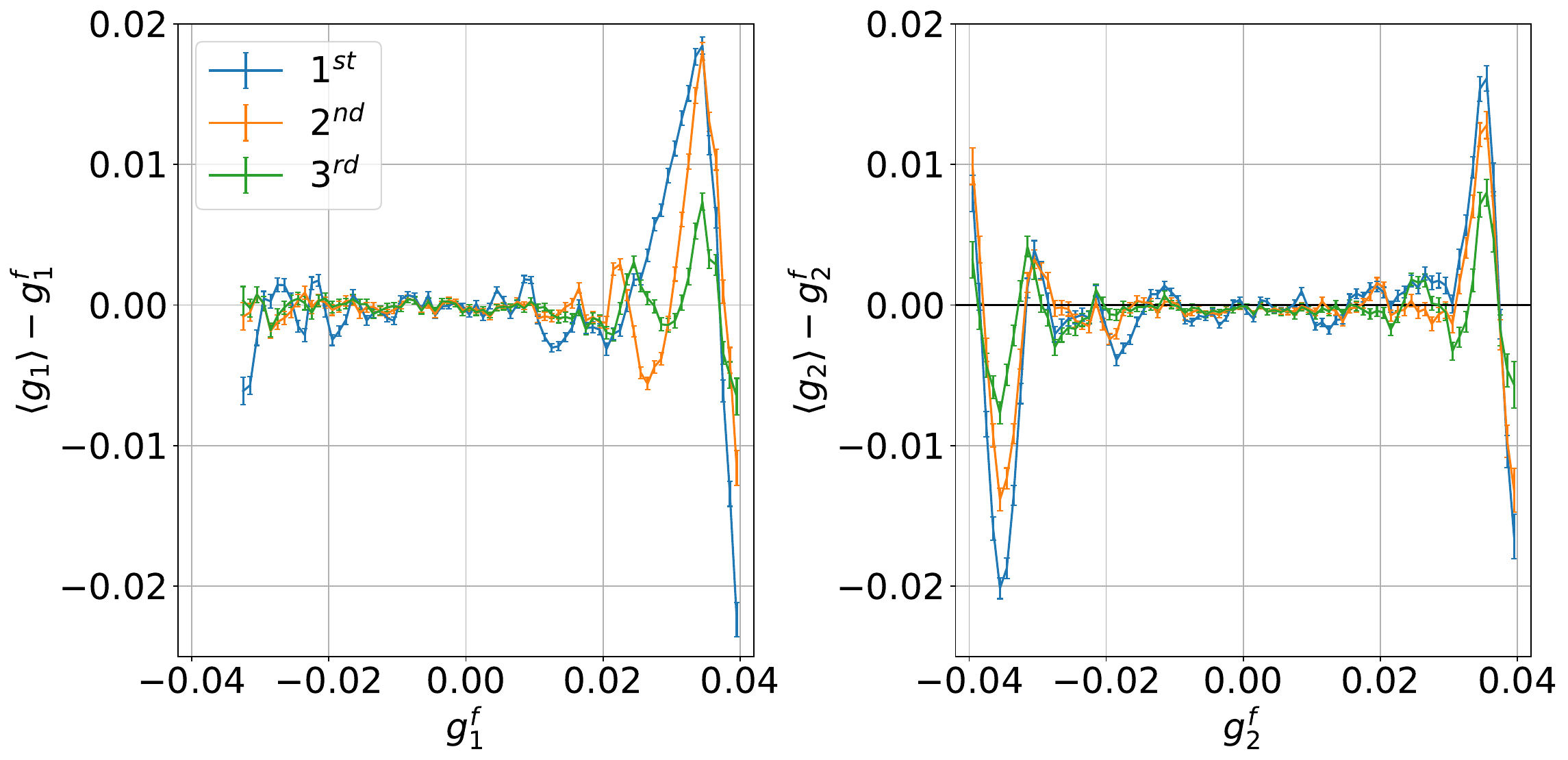}
    \caption{The results of the field distortion tests with different PSF polynomial order. The left and right panels show the residuals of $g_1$ and $g_2$ respectively. Lines with the blue, orange, and green colors show the results of the 1st, 2nd, and 3rd polynomial order for PSF reconstruction.}
    \label{fig:wiggle}
\end{figure}
\begin{figure}[htb!]

    \centering
    \includegraphics[width=\columnwidth]{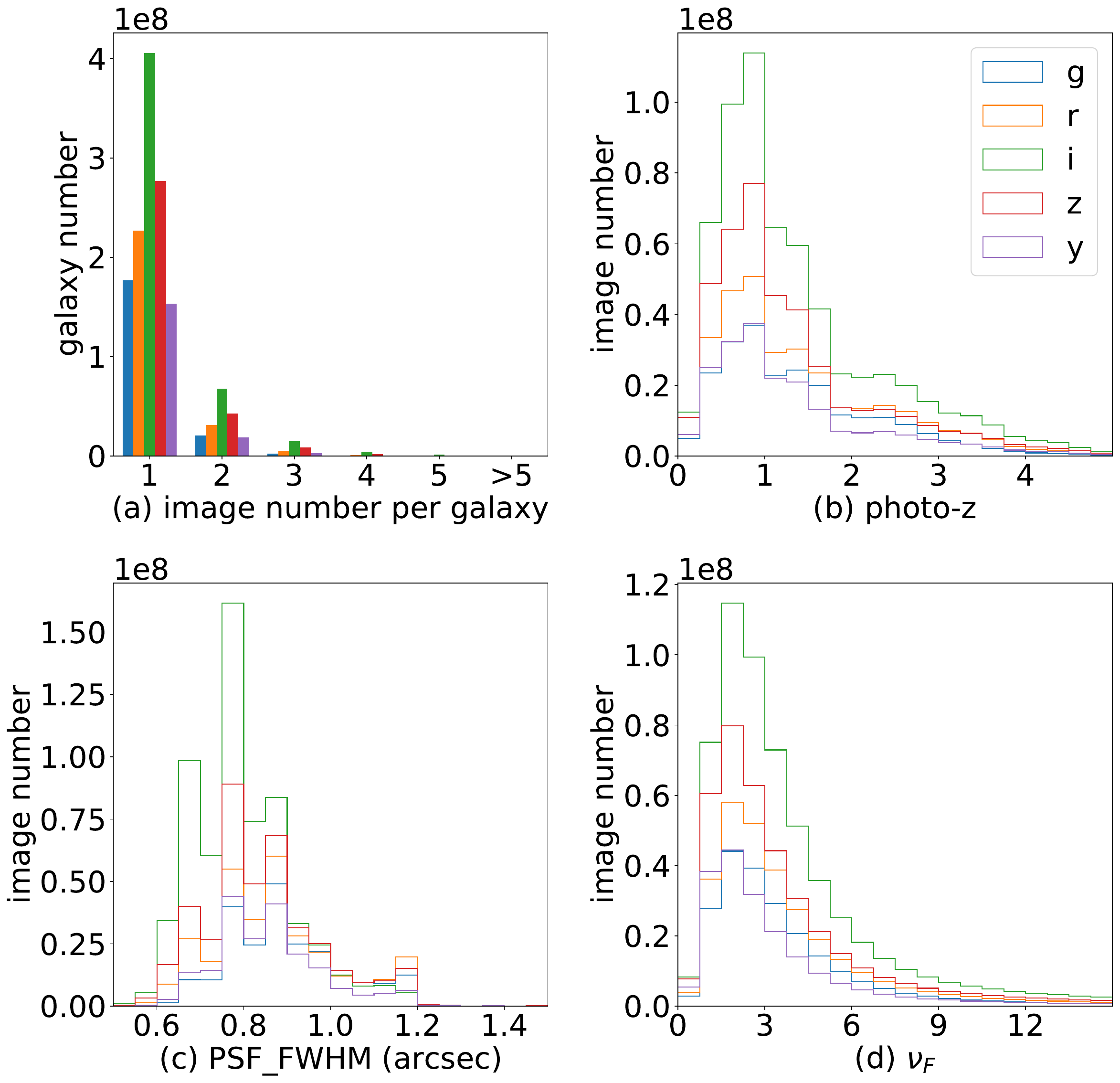}

    \caption{Some statistical properties of our shear catalogs. (a) distribution of image number per galaxy; (b) the photo-z distribution; (c) the FWHM distribution of the PSF; (d) The $v_F$ distribution. Results of different bands are shown with individual colors in all panels.}
    \label{fig:dist}
\end{figure}

In fig.\ref{fig:dist}, we show several statistical properties of our shear catalog. Note that our image processing pipeline treats each exposure independently, each galaxy can yield more than one set of shear estimators, i.e., one set from each valid image. Fig.\ref{fig:dist} (a) shows the distribution of image number per galaxy in each band. Panel (b) of the same figure shows the photo-z distribution of the shear catalog. These redshift information are directly obtained from the source catalog discribed in \ref{HSC_dataset}. Panel (c) and (d) of fig.\ref{fig:dist} show the distributions of $\nu_F$ of the galaxies and the FWHM of the PSF respectively.

To use the shear catalogs, one should apply several cuts to avoid systematic biases. Here we summarize the necessary cuts:
\begin{itemize}
    \item $\vert g^f \vert\le 0.02$, to remove the galaxies near the edges of the focal plane, where the PSF models have large biases.
    \item Remove the galaxies from the CCDs with chip id of 0, 33, 43, 94 due to bad CCD channels.
    \item $\nu_F>5$ to avoid detection-related biases at the faint end.
\end{itemize}

\begin{figure}[htb!]
    \centering

    \includegraphics[width=0.9\columnwidth]{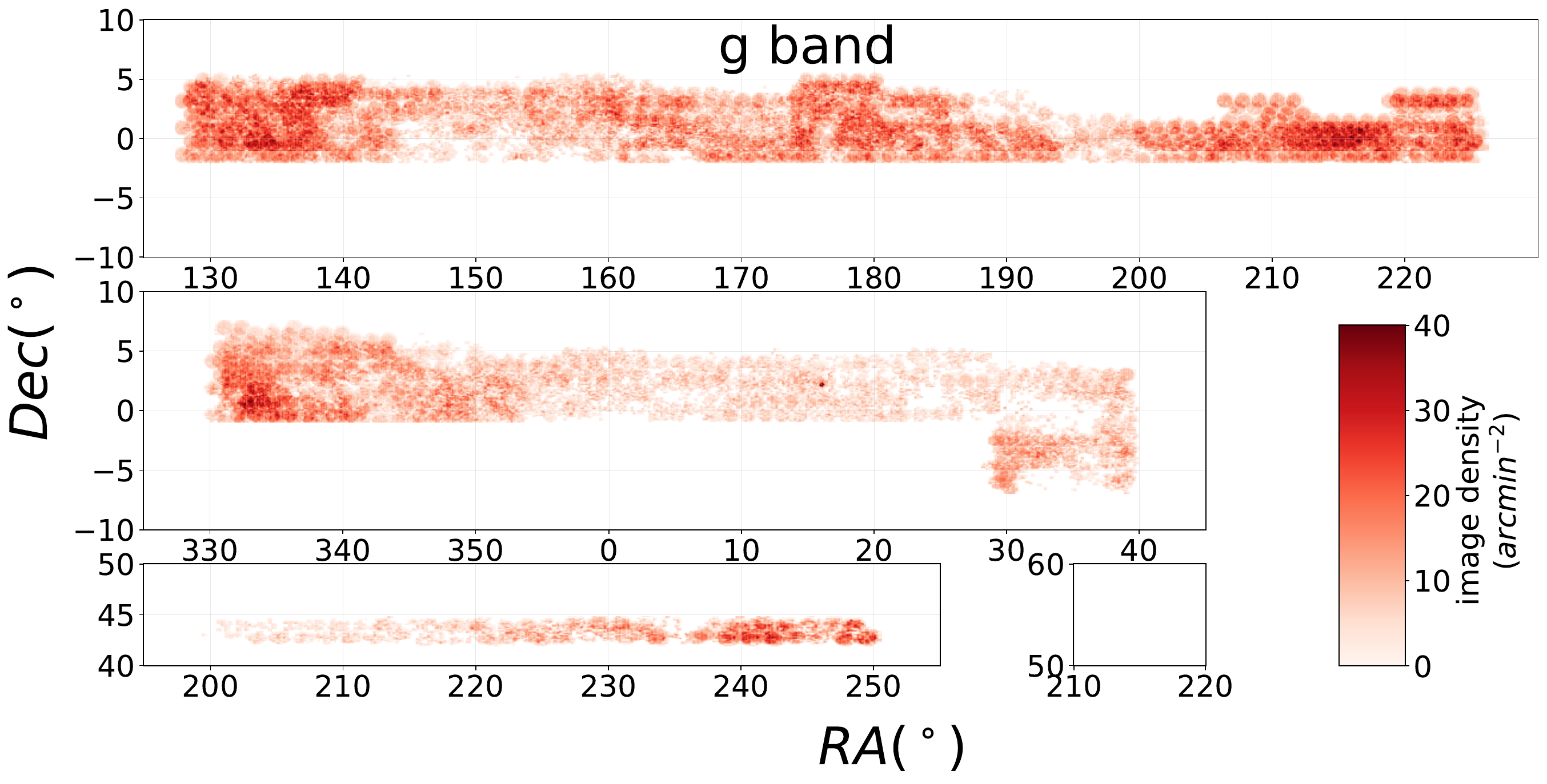} \\
    \includegraphics[width=0.9\columnwidth]{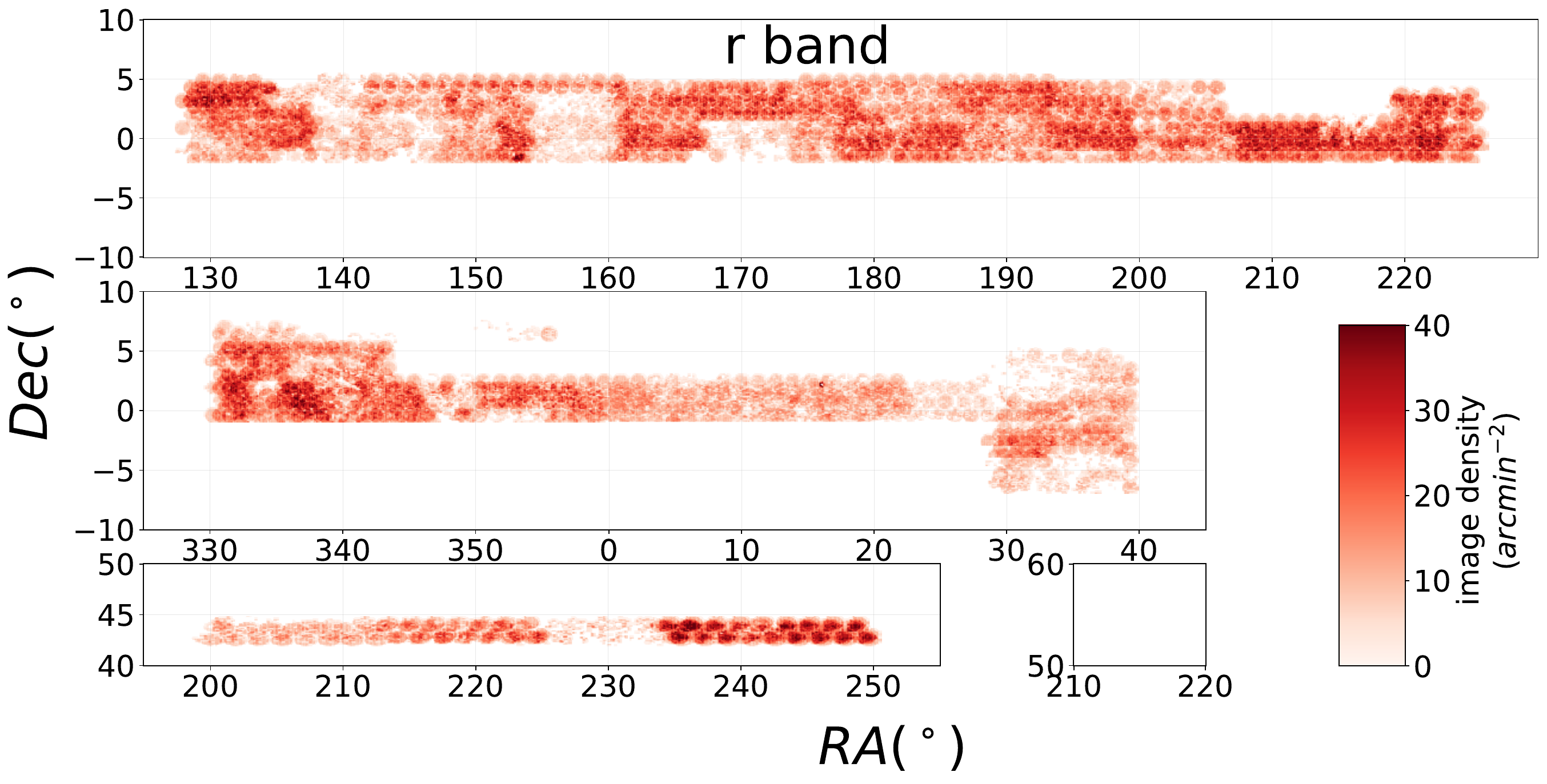} \\
    \includegraphics[width=0.9\columnwidth]{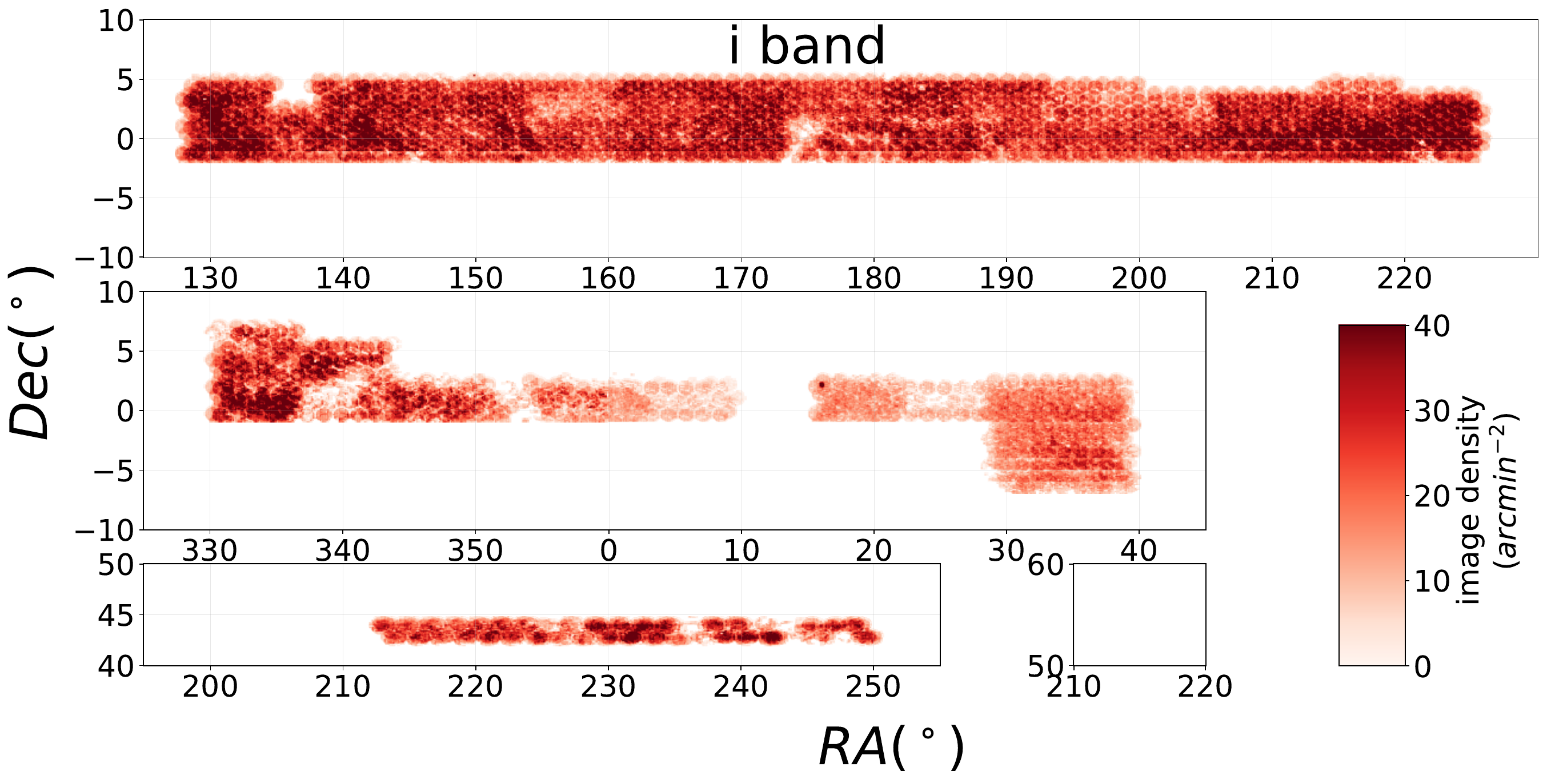} \\
    \includegraphics[width=0.9\columnwidth]{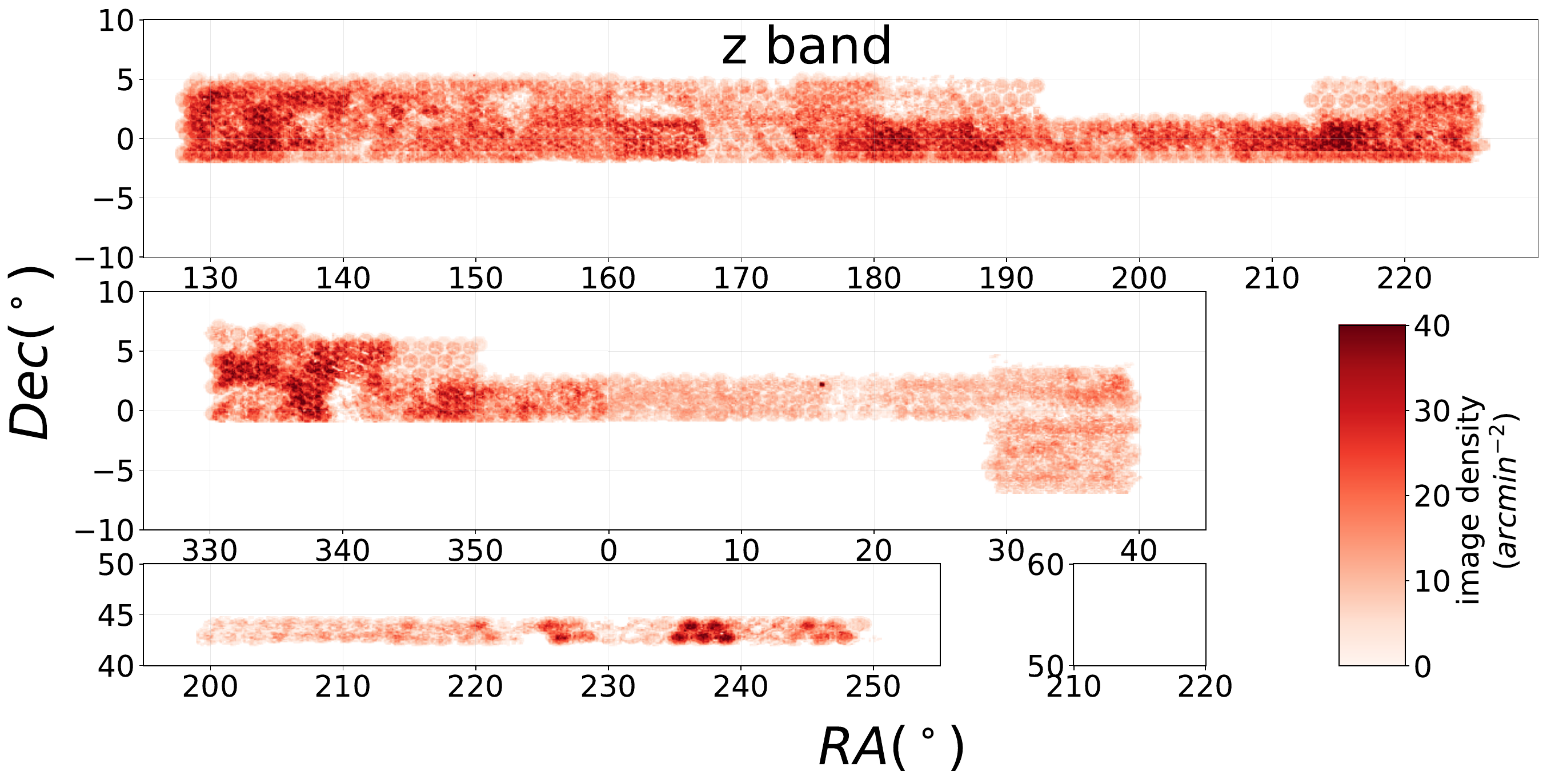} \\
    \includegraphics[width=0.9\columnwidth]{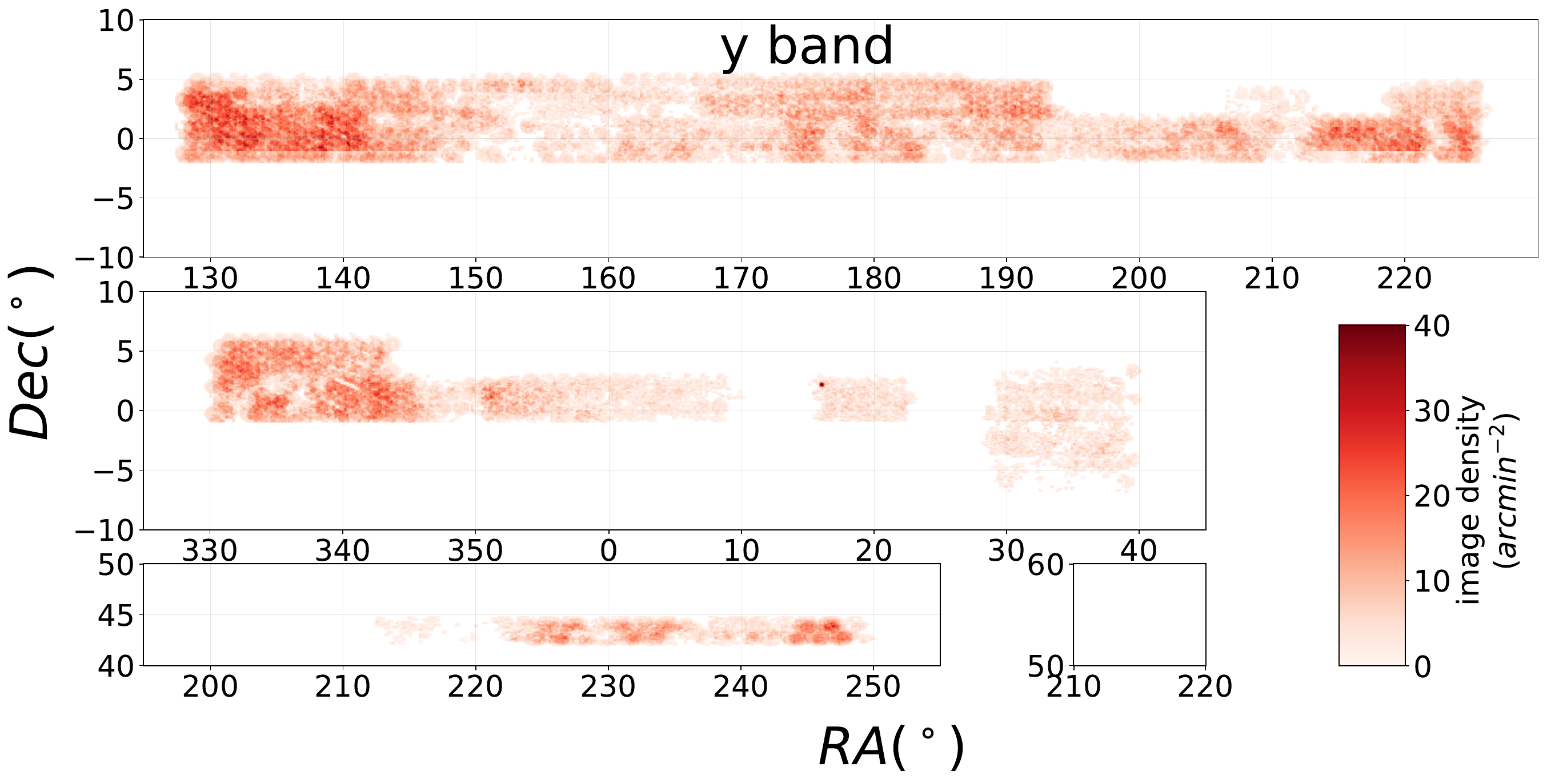}

    \caption{The sky coverages of the shear catalogs in g/r/i/z/y bands after the cuts.}
    \label{fig:footprint}
\end{figure}

In table \ref{tab:catinfo}, we summarize the sky coverage, image density, galaxy density and shape noise per galaxy of the shear catalogs after applying the cuts. All five bands have more than $1200 \deg^2$ sky coverage. The spatial distribution of the galaxy density in each band is shown in fig.\ref{fig:footprint}. Among these five bands, the i-band has the highest galaxy density and lowest shape noise, the r and z bands are a little worse, and followed by the g and y bands. This suggests that the i-band has the best performance, which is understandable considering the observational condition. 
In the next section, we further discuss the performance of each band with galaxy-galaxy lensing. 

\begin{table}[htb!]
    \centering
    \begin{tabular}{ccccc}
        \hline
        \hline
        band&sky&image&galaxy&shape \\
            &coverage&density&density&noise \\ 
            &$({\rm deg^2})$&$({\rm arcmin^{-2}})$&$({\rm arcmin^{-2}})$& \\
        \hline
        g&1343& 8.33& 7.68&0.415 \\
        r&1324&11.60&10.50&0.385 \\
        i&1314&22.28&19.33&0.359 \\
        z&1350&13.84&11.93&0.365 \\
        y&1220& 6.40& 5.66&0.412 \\
        \hline
    \end{tabular}
    \caption{Some basic information of the shear catalogs after the cut, including the sky coverage, the average image/galaxy number density, and the shape noise per galaxy.}
    \label{tab:catinfo}
\end{table}

\section{Multi-Band Performance}
\label{performance}

The field distortion test results in fig.\ref{fig:fd_mc} demonstrate that we have obtained low bias shear catalogs for all five bands from HSCpdr3 datasets. In this section, we use galaxy-galaxy lensing to further test the performance of our multi-band shear catalogs. We compare the results from individual bands to check their consistency, and combine shear catalogs of different bands to look for improvement in accuracy. In addition, we check the consistency between our catalogs and the official HSC year-one shear catalog. 
 
In galaxy-galaxy lensing, the excess surface density $\Delta\Sigma$ (ESD) of the foreground lens is related to the tangential shear of the background galaxy ($\gamma_t$) via a simple formula:  
\begin{equation}
    \begin{aligned}        \Sigma_c\gamma_t(r)&=\Delta\Sigma(r)=\bar{\Sigma}(<r)-\Sigma(r)
    \end{aligned}
\end{equation}
where $\bar{\Sigma}(<r)$ is the average projected surface density within radius r, and $\Sigma(r)$ is the surface density on radius r.
$\Sigma_c=c^2/(4\pi G)\cdot D_{s}/(D_{ls}D_l)$ is the critical surface density, in which $D_s,D_l,D_{ls}$ refer to the angular diameter distances of the source, the lens, and the distance between the source and the lens respectively.

To recover $\Delta\Sigma$ with the FQ shear estimators and the PDF-SYM method, we need to assign its assumed value as $\widehat{\Delta\Sigma}$, and change it so that the PDF of the modified shear estimator $\hat{G}_{\mathrm{t}}$ is best symmetrized. The modified shear estimator is defined as:
\begin{equation}
    \hat{G}_{\mathrm{t}}=G_{\mathrm{t}}\left(z_{\mathrm{s}}\right)-\frac{\widehat{\Delta \Sigma}}{\Sigma_c\left(z_l, z_s\right)} \cdot\left(N+U_{\mathrm{t}}\right)\left(z_{\mathrm{s}}\right)
\end{equation}
where $z_{\mathrm{l}},z_{\mathrm{s}}$ stand for the lens and source redshift, while $G_{\mathrm{t}}, U_{\mathrm{t}}$ are defined in the tangential direction. More details of galaxy-galaxy lensing with FQ shear estimators are discussed in \cite{Jiaqi_ggl}. 

We choose the LOWZ and CMASS galaxies from Baryon Oscillation Spectroscopic Survey (BOSS \cite{boss}) as the foreground lens. Each group are divided into two subsamples (labeled as L1, L2, C1, C2) according to their redshift. The redshfit range of these four samples are [0.15-0.31],[0.31-0.43],[0.43,0.54],[0.54-0.70]. The measurement is carried out within the overlapped regions of the lens sample and our shear catalogs. The sky coverage is almost the same as that of our shear catalogs.

\begin{figure*}[htb!]
    \centering
    \includegraphics[width=0.95\textwidth]{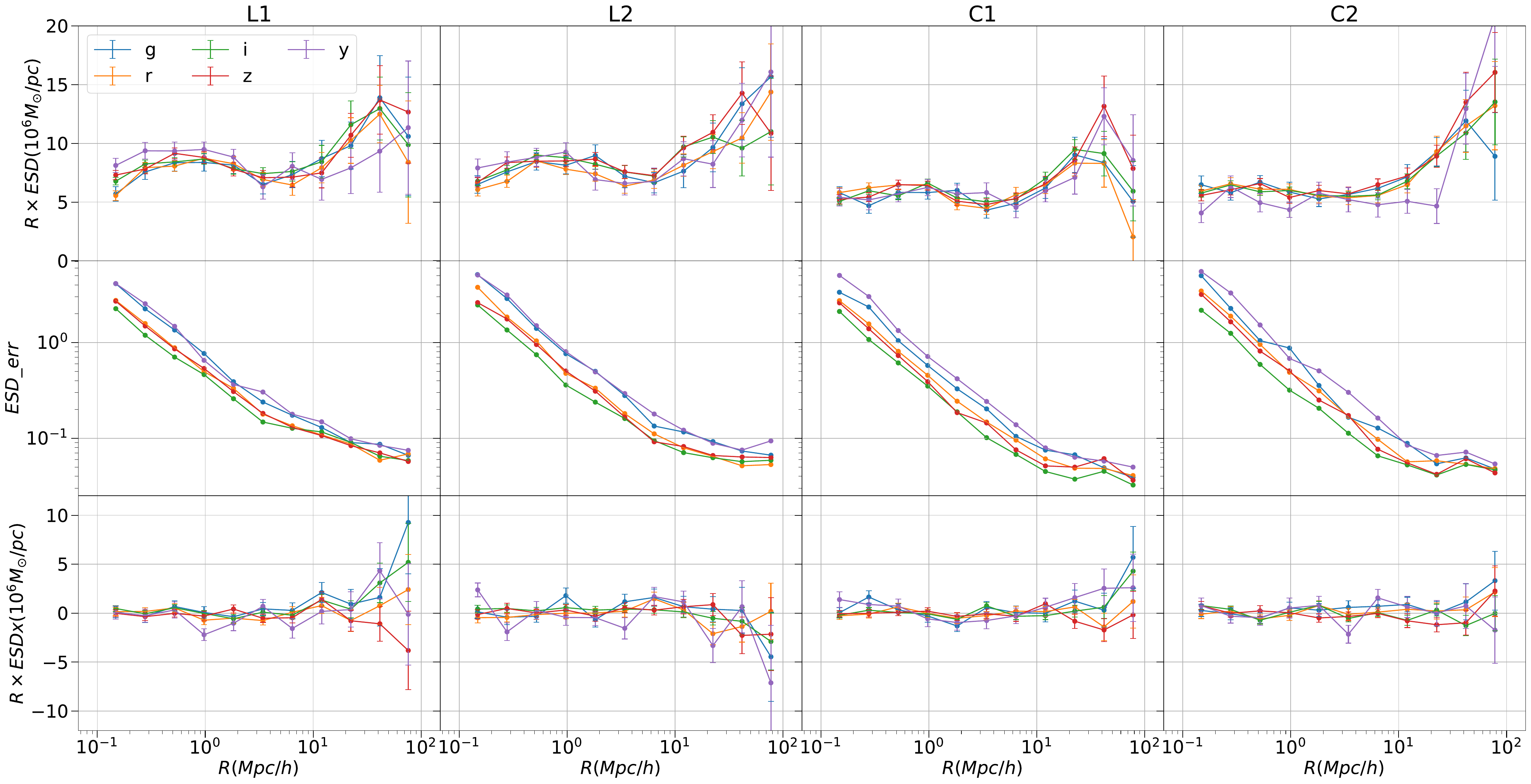}
    \caption{Galaxy-galaxy lensing results. Each column shows results from difference lens samples: L1, L2, C1, C2 from left to right. The upper, middle, and bottom rows show the results of ESD, error of ESD, and ESD\_x respectively. Results from difference bands are labeled with difference colors in every panels. The error bars are estimated with 100 jackknife subsamples. And all the data points are subtracted by the results from the random sample (10 times larger). }
    \label{fig:ggl_all}
\end{figure*}

In fig.\ref{fig:ggl_all}, we show in the upper, middle, and lower panels the measured R*ESD, ESD error (estimated by 100 jackknife samples) and ESD\_x (averaged cross shear) respectively. All the data points are subtracted by the results from the random sample (10 times larger). The first row of the figure shows that the five bands are consistent with each other in every lens sample. These results demonstrate the robustness of our shear catalogs. From the second row, we can compare the performance of each band. The plots show that the i-band yields the lowest error, followed by the r \& z bands, and then by the g \& y. This is consistent with the conclusion we draw from galaxy density and shape noise. And for the last rows, ESD\_x, we can see that these B modes are consistent with zero, which again indicates that our shear catalogs have insignificant systematic biases. 

\begin{figure*}[htb!]
    \centering
    \includegraphics[width=0.95\textwidth]{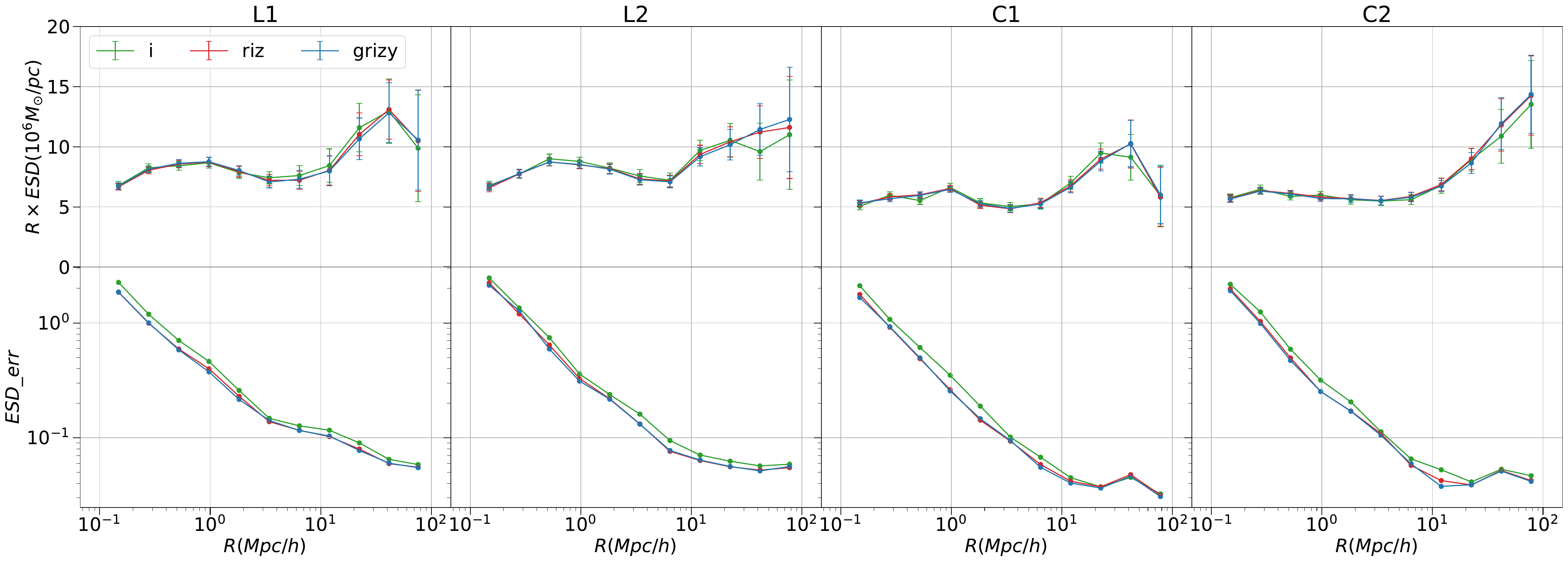}
    \caption{Similar to Figure \ref{fig:ggl_all}, but without the bottom panels for ESD\_x. Results from combining the r/i/z bands and all five bands are added. }
    \label{fig:ggl_riz}
\end{figure*}

We further show in fig.\ref{fig:ggl_riz} the g-g lensing results using the combined shear catalogs of the r, i, and z bands, as well as the combination of all five bands. In the combined shear catalog, we simply treat shear estimators from different bands as individual measurement. The results show that combining five bands yields even lower errors compared to those from individual bands. On average, the error decreases by 14.7\%, 14.0\%, 16.2\%, 17.3\% with respect to the i-band-only case for the L1, L2, C1, C2 lens samples respectively. This shows that we indeed gain extra information by combining measurements from different bands.

\begin{figure*}[htb!]
    \centering
    \includegraphics[width=0.95\textwidth]{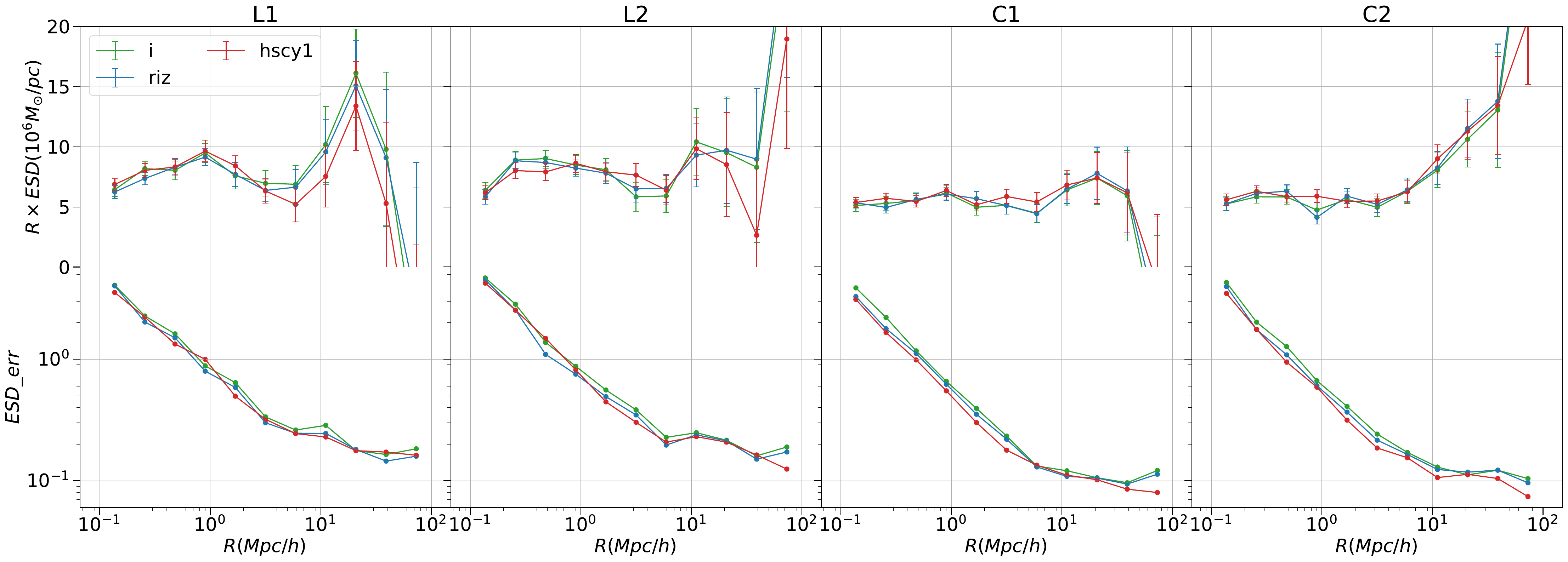}
    \caption{Similar to fig.\ref{fig:ggl_all}. Results from the official HSC year-one catalog are added.}
    \label{fig:ggl_y1}
\end{figure*}

Finally, we use galaxy-galaxy lensing to check the consistency between our shear catalog and the official HSC shear catalog.\cite{HSCY1shear} reported the HSC year-one i-band shear catalog measured with the re-Gaussianization method \citep{regauss}. In this measurement we only use the overlapped regions of the foreground lens, the HSC year-one official catalog, and our catalogs. The results are shown in fig.\ref{fig:ggl_y1}.As indicated in the figure, our results from four lens sample are consistent with those from the official catalog. The errors from the i-band only case are slightly larger than those from the official catalog, but the combined catalog of r, i, and z shows comparable uncertainties in the measurement with respect to the official catalog.

\section{Conclusion \& Discussions}
\label{conclusion}

Ongoing large area photometric surveys usually cover several broad bands in their wide layer. Typically, shape measurement is carried out within certain band(s) with better observing conditions. Given the variations of the image qualities in different bands, we are curious about the robustness of each band for shear measurement, and how much we can improve the shear statistics by combining the multi-band shear catalogs.

In this paper, we make use of the HSCpdr3 dataset, which contains five (g,r,i,z,y) broad bands, to test its multi-band performance. For shear measurement, we adopt the Fourier\_Quad method and its image processing pipeline. The pipeline was first introduced in \cite{FQ_fd} for processing the CFHTLenS data, and then in \cite{FQ_decals} for the DECaLS data. Modifications of the pipeline used in this work are given in \S\ref{image_processing}.
 
Using the field distortion test, we demonstrate in fig.\ref{fig:fd_mc} that our shear catalogs of different bands all have very high qualities. In most of the panels, we see that the multiplicative biases are consistent with zero within about $2\sigma$. Only the $g_2$ results in the r and z bands have about somewhat significant 2-3\% multiplicative biases, which may deserve some further studies. The additive biases are also mostly minor. The $g_1$ results of y-band has a particularly large additive bias, which is likely due to the scatter light problem \cite{hscdr2}. 

We note that there are some issues left in our image processing. For example, our PSF models still have large residuals at outskirts of focal plane. Through the focal plane null test in \S\ref{null_test}, we show that these residuals directly cause the shear errors, which make us remove these regions in our final shear catalogs. As these structures tend to be stable patterns on the focal plane, we can model them using all the star candidates from survey. This work is reported in \cite{alonso2024}. Another thing is about the deblending effect, which is significant for the next generation surveys \citep{Arcelin2020,LiuDZ_2023}. Currently we simply throw away blended galaxies with redshift difference larger than 0.1, or treat them as a single source if their redshift distance is small. More refined treatment will be considered in our future work.

We further use galaxy-galaxy lensing to check the performance of our multi-band shear catalogs. The ESD measurement with individual band shows that all five bands are consistent with each other. Considering the galaxy density, shape noise and SNR of ESD, we conclude that i-band have the best performance, followed by r,z bands, and then g,y bands. We combine r,i,z band to measure ESD again and get $\sim 15\%$ lower errorbars than the i-band-only result. This suggests that multi-band shear measurement has improvement in shear statistics. In addition, we compare our shear catalogs with the HSC year-one official shear catalogs. The results are all consistence with each other.

\begin{figure}[htb!]

    \centering
    \includegraphics[width=\columnwidth]{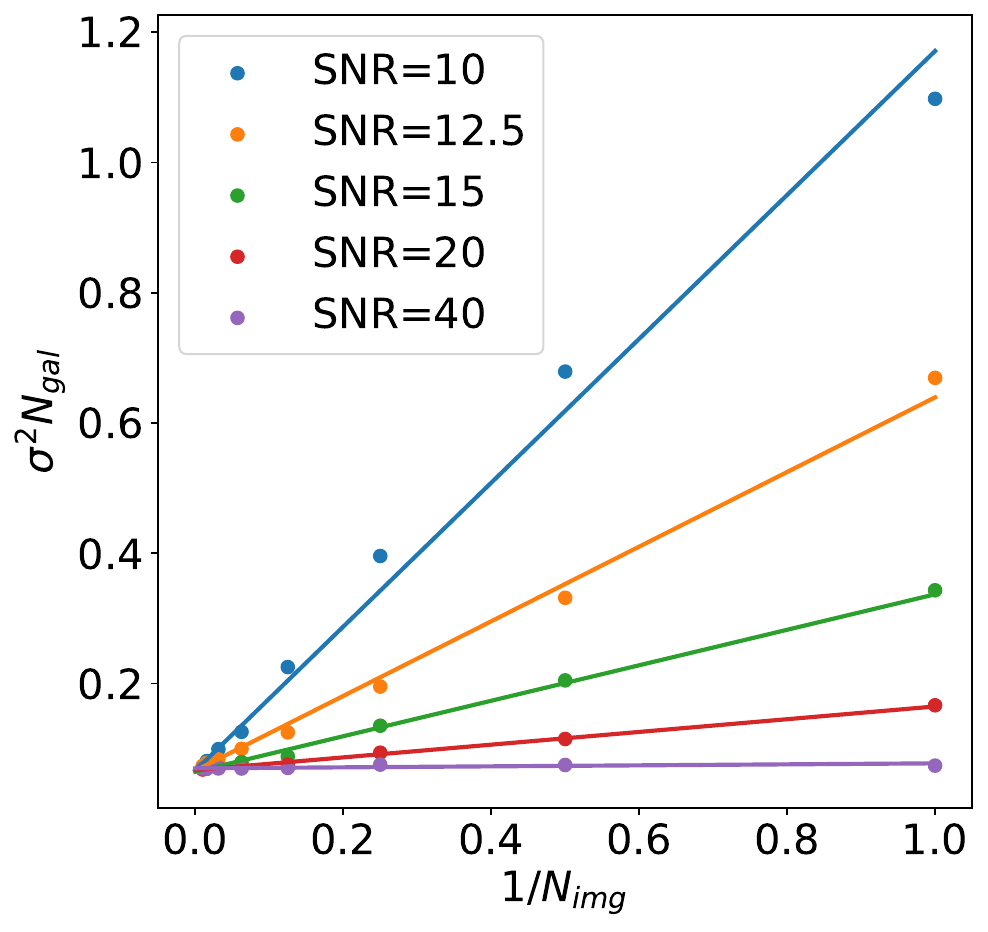}

    \caption{Change of the shape noise with the galaxy image/exposure number in FQ. Dots with different colors show the results from simulations with galaxies of different SNR. The lines are the best-fit linear relations between the square of the shape noise and the inverse of the image number.}
    \label{fig:sigma_Nimg}
\end{figure}

As for multi-band shear statistics, we find that the SNR of galaxy-galaxy lensing does rise with the increasing number of bands. This enhancement mainly comes from the cancellation of the background noise. It should ultimately be limited by the galaxy shape noise. Assuming that both the background noise and shape noise act like Poisson noise to the shear estimators, we expect the following relation to hold:
\begin{equation}
\label{noise_scaling}
    \sigma^2N_{gal} = \sigma_{sp}^2 + \sigma_{bk}^2/N_{img},
\end{equation}
in which $N_{gal}$ is the number of distinct galaxies, and $N_{img}$ is the number of exposures of each galaxy, which can come from different bands. $\sigma$ is the overall uncertainty of the shear recovery. $\sigma_{sp}$ is the shape noise (per galaxy), and  $\sigma_{bk}$ is the average background noise (for a given SNR) per galaxy image. As there are not so many exposures/images for each galaxy in real data, we carry out a simple image simulation using Galsim \citep{galsim} to study this issue. We generate $10^5$ galaxies with the De Vaucouleurs profile. All the galaxies are distorted by a constant shear and convolved with a Gaussian PSF. For each galaxy, we keep a certain SNR and draw 100 images with different Gaussian random noise. We estimate the shear using different numbers of images, and measure the uncertainty of the recovered shear as a function of the image number. The results are shown in fig.\ref{fig:sigma_Nimg}. It confirms the formula given in eq.\ref{noise_scaling}. It is clear that there is room for reducing the measurement error by increasing the number of exposures, especially for the faint sources.

In fig.\ref{fig:sigma_Nimg}, we learn that for faint galaxies, which are dominated by the background noise, the measurement from different images (bands) can significantly enhance the shear statistics. But for the bright sources (SNR $\gtrsim$ 20), multi-image measurement only results in a small reduction of the error. High redshift galaxies tend to be fainter than low redshift ones, but carry greater lensing signal. The advantage of multi-band shear measurement may be more obvious in tomographic shear-shear correlation, which we will report in another work using the shear catalogs produced in this paper.

Meanwhile, we find that the direct stacking of the FQ shear estimators (or image power spectrum) is not as efficient as stacking real images. As FQ shear estimators use the image power spectrum, the SNR in Fourier space is proportional to the square of the SNR in real space. Therefore, for each galaxy, to achieve the same noise reduction of real space stacking with $N$ images, we need $N^2$ images in FQ. This fact can be observed in fig.\ref{fig:sigma_Nimg}. Thus treating FQ shear estimators from different images (or bands) as individual measurement is not yet the optimal way in multi-band shear statistics. We will try to develop new ways of combining multi-band/exposure shear estimators of FQ in a future work. 

Finally, very recently, in our another work (Shen et al., in preparation), we find that significant multiplicative biases can arise when the source galaxies are binned according to their redshifts. This is likely caused by a strong selection effect, the origin of which is still under our investigation. Fortunately, all of our shear catalogs contain the field distortion information. These selection-based biases can be estimated onsite whenever necessary.  

We will make our shear catalogs publicly available upon the publication of this work \footnote{\url{https://gax.sjtu.edu.cn/data/DESI.html}}. The image process pipeline and intermediate products are also available by request.

\section{acknowledgements}

This work is supported by the National Key Basic Research and Development Program of China (2023YFA1607800, 2023YFA1607802), the NSFC grants (11621303, 11890691, 12073017), and the science research grants from China Manned Space Project (No. CMS-CSST-2021-A01). WW is supported by NSFC grants (12022307,12273021). The computations in this paper were run on the $\pi$ 2.0 cluster supported by the Center of High Performance Computing at Shanghai Jiaotong University, and the Gravity supercomputer of the Astronomy Department, Shanghai Jiaotong University.

The Hyper Suprime-Cam (HSC) collaboration includes the astronomical communities of Japan and Taiwan, and Princeton University. The HSC instrumentation and software were developed by the National Astronomical Observatory of Japan (NAOJ), the Kavli Institute for the Physics and Mathematics of the Universe (Kavli IPMU), the University of Tokyo, the High Energy Accelerator Research Organization (KEK), the Academia Sinica Institute for Astronomy and Astrophysics in Taiwan (ASIAA), and Princeton University. Funding was contributed by the FIRST program from the Japanese Cabinet Office, the Ministry of Education, Culture, Sports, Science and Technology (MEXT), the Japan Society for the Promotion of Science (JSPS), Japan Science and Technology Agency (JST), the Toray Science Foundation, NAOJ, Kavli IPMU, KEK, ASIAA, and Princeton University. 

This paper makes use of software developed for Vera C. Rubin Observatory. We thank the Rubin Observatory for making their code available as free software at http://pipelines.lsst.io/.

This paper is based on data collected at the Subaru Telescope and retrieved from the HSC data archive system, which is operated by the Subaru Telescope and Astronomy Data Center (ADC) at NAOJ. Data analysis was in part carried out with the cooperation of Center for Computational Astrophysics (CfCA), NAOJ. We are honored and grateful for the opportunity of observing the Universe from Maunakea, which has the cultural, historical and natural significance in Hawaii. 

\bibliography{thesis}{}

\begin{thebibliography}{}
\expandafter\ifx\csname natexlab\endcsname\relax\def\natexlab#1{#1}\fi
\providecommand{\url}[1]{\href{#1}{#1}}
\providecommand{\dodoi}[1]{doi:~\href{http://doi.org/#1}{\nolinkurl{#1}}}
\providecommand{\doeprint}[1]{\href{http://ascl.net/#1}{\nolinkurl{http://ascl.net/#1}}}
\providecommand{\doarXiv}[1]{\href{https://arxiv.org/abs/#1}{\nolinkurl{https://arxiv.org/abs/#1}}}

\bibitem[{{Abbott} {et~al.}(2022){Abbott}, {Aguena}, {Alarcon}, {Allam},
  {Alves}, {Amon}, {Andrade-Oliveira}, {Annis}, {Avila}, {Bacon}, {Baxter},
  {Bechtol}, {Becker}, {Bernstein}, {Bhargava}, {Birrer}, {Blazek},
  {Brandao-Souza}, {Bridle}, {Brooks}, {Buckley-Geer}, {Burke}, {Camacho},
  {Campos}, {Carnero Rosell}, {Carrasco Kind}, {Carretero}, {Castander},
  {Cawthon}, {Chang}, {Chen}, {Chen}, {Choi}, {Conselice}, {Cordero},
  {Costanzi}, {Crocce}, {da Costa}, {da Silva Pereira}, {Davis}, {Davis}, {De
  Vicente}, {DeRose}, {Desai}, {Di Valentino}, {Diehl}, {Dietrich}, {Dodelson},
  {Doel}, {Doux}, {Drlica-Wagner}, {Eckert}, {Eifler}, {Elsner}, {Elvin-Poole},
  {Everett}, {Evrard}, {Fang}, {Farahi}, {Fernandez}, {Ferrero}, {Fert{\'e}},
  {Fosalba}, {Friedrich}, {Frieman}, {Garc{\'\i}a-Bellido}, {Gatti},
  {Gaztanaga}, {Gerdes}, {Giannantonio}, {Giannini}, {Gruen}, {Gruendl},
  {Gschwend}, {Gutierrez}, {Harrison}, {Hartley}, {Herner}, {Hinton},
  {Hollowood}, {Honscheid}, {Hoyle}, {Huff}, {Huterer}, {Jain}, {James},
  {Jarvis}, {Jeffrey}, {Jeltema}, {Kovacs}, {Krause}, {Kron}, {Kuehn},
  {Kuropatkin}, {Lahav}, {Leget}, {Lemos}, {Liddle}, {Lidman}, {Lima}, {Lin},
  {MacCrann}, {Maia}, {Marshall}, {Martini}, {McCullough}, {Melchior},
  {Mena-Fern{\'a}ndez}, {Menanteau}, {Miquel}, {Mohr}, {Morgan}, {Muir},
  {Myles}, {Nadathur}, {Navarro-Alsina}, {Nichol}, {Ogando}, {Omori},
  {Palmese}, {Pandey}, {Park}, {Paz-Chinch{\'o}n}, {Petravick}, {Pieres},
  {Plazas Malag{\'o}n}, {Porredon}, {Prat}, {Raveri}, {Rodriguez-Monroy},
  {Rollins}, {Romer}, {Roodman}, {Rosenfeld}, {Ross}, {Rykoff}, {Samuroff},
  {S{\'a}nchez}, {Sanchez}, {Sanchez}, {Sanchez Cid}, {Scarpine}, {Schubnell},
  {Scolnic}, {Secco}, {Serrano}, {Sevilla-Noarbe}, {Sheldon}, {Shin}, {Smith},
  {Soares-Santos}, {Suchyta}, {Swanson}, {Tabbutt}, {Tarle}, {Thomas}, {To},
  {Troja}, {Troxel}, {Tucker}, {Tutusaus}, {Varga}, {Walker}, {Weaverdyck},
  {Wechsler}, {Weller}, {Yanny}, {Yin}, {Zhang}, {Zuntz}, \& {DES
  Collaboration}}]{DESY3cosmos}
{Abbott}, T.~M.~C., {Aguena}, M., {Alarcon}, A., {et~al.} 2022, prd, 105,
  023520, \dodoi{10.1103/PhysRevD.105.023520}

\bibitem[{{Aihara} {et~al.}(2018){Aihara}, {Arimoto}, {Armstrong}, {Arnouts},
  {Bahcall}, {Bickerton}, {Bosch}, {Bundy}, {Capak}, {Chan}, {Chiba}, {Coupon},
  {Egami}, {Enoki}, {Finet}, {Fujimori}, {Fujimoto}, {Furusawa}, {Furusawa},
  {Goto}, {Goulding}, {Greco}, {Greene}, {Gunn}, {Hamana}, {Harikane},
  {Hashimoto}, {Hattori}, {Hayashi}, {Hayashi}, {He{\l}miniak}, {Higuchi},
  {Hikage}, {Ho}, {Hsieh}, {Huang}, {Huang}, {Ikeda}, {Imanishi}, {Inoue},
  {Iwasawa}, {Iwata}, {Jaelani}, {Jian}, {Kamata}, {Karoji}, {Kashikawa},
  {Katayama}, {Kawanomoto}, {Kayo}, {Koda}, {Koike}, {Kojima}, {Komiyama},
  {Konno}, {Koshida}, {Koyama}, {Kusakabe}, {Leauthaud}, {Lee}, {Lin}, {Lin},
  {Lupton}, {Mandelbaum}, {Matsuoka}, {Medezinski}, {Mineo}, {Miyama},
  {Miyatake}, {Miyazaki}, {Momose}, {More}, {More}, {Moritani}, {Moriya},
  {Morokuma}, {Mukae}, {Murata}, {Murayama}, {Nagao}, {Nakata}, {Niida},
  {Niikura}, {Nishizawa}, {Obuchi}, {Oguri}, {Oishi}, {Okabe}, {Okamoto},
  {Okura}, {Ono}, {Onodera}, {Onoue}, {Osato}, {Ouchi}, {Price}, {Pyo}, {Sako},
  {Sawicki}, {Shibuya}, {Shimasaku}, {Shimono}, {Shirasaki}, {Silverman},
  {Simet}, {Speagle}, {Spergel}, {Strauss}, {Sugahara}, {Sugiyama}, {Suto},
  {Suyu}, {Suzuki}, {Tait}, {Takada}, {Takata}, {Tamura}, {Tanaka}, {Tanaka},
  {Tanaka}, {Tanaka}, {Terai}, {Terashima}, {Toba}, {Tominaga}, {Toshikawa},
  {Turner}, {Uchida}, {Uchiyama}, {Umetsu}, {Uraguchi}, {Urata}, {Usuda},
  {Utsumi}, {Wang}, {Wang}, {Wong}, {Yabe}, {Yamada}, {Yamanoi}, {Yasuda},
  {Yeh}, {Yonehara}, \& {Yuma}}]{HSCsurvey}
{Aihara}, H., {Arimoto}, N., {Armstrong}, R., {et~al.} 2018, \pasj, 70, S4,
  \dodoi{10.1093/pasj/psx066}

\bibitem[{{Aihara} {et~al.}(2019){Aihara}, {AlSayyad}, {Ando}, {Armstrong},
  {Bosch}, {Egami}, {Furusawa}, {Furusawa}, {Goulding}, {Harikane}, {Hikage},
  {Ho}, {Hsieh}, {Huang}, {Ikeda}, {Imanishi}, {Ito}, {Iwata}, {Jaelani},
  {Kakuma}, {Kawana}, {Kikuta}, {Kobayashi}, {Koike}, {Komiyama}, {Li},
  {Liang}, {Lin}, {Luo}, {Lupton}, {Lust}, {MacArthur}, {Matsuoka}, {Mineo},
  {Miyatake}, {Miyazaki}, {More}, {Murata}, {Namiki}, {Nishizawa}, {Oguri},
  {Okabe}, {Okamoto}, {Okura}, {Ono}, {Onodera}, {Onoue}, {Osato}, {Ouchi},
  {Shibuya}, {Strauss}, {Sugiyama}, {Suto}, {Takada}, {Takagi}, {Takata},
  {Takita}, {Tanaka}, {Terai}, {Toba}, {Uchiyama}, {Utsumi}, {Wang}, {Wang}, \&
  {Yamada}}]{hscdr2}
{Aihara}, H., {AlSayyad}, Y., {Ando}, M., {et~al.} 2019, \pasj, 71, 114,
  \dodoi{10.1093/pasj/psz103}

\bibitem[{{Aihara} {et~al.}(2022){Aihara}, {AlSayyad}, {Ando}, {Armstrong},
  {Bosch}, {Egami}, {Furusawa}, {Furusawa}, {Harasawa}, {Harikane}, {Hsieh},
  {Ikeda}, {Ito}, {Iwata}, {Kodama}, {Koike}, {Kokubo}, {Komiyama}, {Li},
  {Liang}, {Lin}, {Lupton}, {Lust}, {MacArthur}, {Mawatari}, {Mineo},
  {Miyatake}, {Miyazaki}, {More}, {Morishima}, {Murayama}, {Nakajima},
  {Nakata}, {Nishizawa}, {Oguri}, {Okabe}, {Okura}, {Ono}, {Osato}, {Ouchi},
  {Pan}, {Plazas Malag{\'o}n}, {Price}, {Reed}, {Rykoff}, {Shibuya},
  {Simunovic}, {Strauss}, {Sugimori}, {Suto}, {Suzuki}, {Takada}, {Takagi},
  {Takata}, {Takita}, {Tanaka}, {Tang}, {Taranu}, {Terai}, {Toba}, {Turner},
  {Uchiyama}, {Vijarnwannaluk}, {Waters}, {Yamada}, {Yamamoto}, \&
  {Yamashita}}]{hscdr3}
---. 2022, \pasj, 74, 247, \dodoi{10.1093/pasj/psab122}

\bibitem[{Alonso {et~al.}(2024)Alonso, Zhang, \& Liu}]{alonso2024}
Alonso, P., Zhang, J., \& Liu, C. 2024, A Hierarchical PSF Reconstruction
  Method.
\newblock \doarXiv{2404.15795}

\bibitem[{{Alonso} {et~al.}(2023){Alonso}, {Wang}, {Zhang}, {Li}, {Shao},
  {Guo}, {He}, {Hao}, \& {Shi}}]{pedro_ggl}
{Alonso}, P., {Wang}, W., {Zhang}, J., {et~al.} 2023, \apj, 947, 19,
  \dodoi{10.3847/1538-4357/acbf4a}

\bibitem[{Arcelin {et~al.}(2020)Arcelin, Doux, Aubourg, Roucelle, \&
  Collaboration)}]{Arcelin2020}
Arcelin, B., Doux, C., Aubourg, E., Roucelle, C., \& Collaboration), T. L. D.
  E.~S. 2020, Monthly Notices of the Royal Astronomical Society, 500, 531,
  \dodoi{10.1093/mnras/staa3062}

\bibitem[{{Asgari} {et~al.}(2021){Asgari}, {Lin}, {Joachimi}, {Giblin},
  {Heymans}, {Hildebrandt}, {Kannawadi}, {St{\"o}lzner}, {Tr{\"o}ster}, {van
  den Busch}, {Wright}, {Bilicki}, {Blake}, {de Jong}, {Dvornik}, {Erben},
  {Getman}, {Hoekstra}, {K{\"o}hlinger}, {Kuijken}, {Miller}, {Radovich},
  {Schneider}, {Shan}, \& {Valentijn}}]{KiDS1000cosmos}
{Asgari}, M., {Lin}, C.-A., {Joachimi}, B., {et~al.} 2021, \aap, 645, A104,
  \dodoi{10.1051/0004-6361/202039070}

\bibitem[{{Bacon} {et~al.}(2000){Bacon}, {Refregier}, \&
  {Ellis}}]{2000MNRAS.318..625B}
{Bacon}, D.~J., {Refregier}, A.~R., \& {Ellis}, R.~S. 2000, \mnras, 318, 625,
  \dodoi{10.1046/j.1365-8711.2000.03851.x}

\bibitem[{{Bosch} {et~al.}(2018){Bosch}, {Armstrong}, {Bickerton}, {Furusawa},
  {Ikeda}, {Koike}, {Lupton}, {Mineo}, {Price}, {Takata}, {Tanaka}, {Yasuda},
  {AlSayyad}, {Becker}, {Coulton}, {Coupon}, {Garmilla}, {Huang}, {Krughoff},
  {Lang}, {Leauthaud}, {Lim}, {Lust}, {MacArthur}, {Mandelbaum}, {Miyatake},
  {Miyazaki}, {Murata}, {More}, {Okura}, {Owen}, {Swinbank}, {Strauss},
  {Yamada}, \& {Yamanoi}}]{hscpipe}
{Bosch}, J., {Armstrong}, R., {Bickerton}, S., {et~al.} 2018, \pasj, 70, S5,
  \dodoi{10.1093/pasj/psx080}

\bibitem[{{Calabretta} \& {Greisen}(2002)}]{wcs2}
{Calabretta}, M.~R., \& {Greisen}, E.~W. 2002, \aap, 395, 1077,
  \dodoi{10.1051/0004-6361:20021327}

\bibitem[{{Dark Energy Survey Collaboration} {et~al.}(2016){Dark Energy Survey
  Collaboration}, {Abbott}, {Abdalla}, {Aleksi{\'c}}, {Allam}, {Amara},
  {Bacon}, {Balbinot}, {Banerji}, {Bechtol}, {Benoit-L{\'e}vy}, {Bernstein},
  {Bertin}, {Blazek}, {Bonnett}, {Bridle}, {Brooks}, {Brunner}, {Buckley-Geer},
  {Burke}, {Caminha}, {Capozzi}, {Carlsen}, {Carnero-Rosell}, {Carollo},
  {Carrasco-Kind}, {Carretero}, {Castander}, {Clerkin}, {Collett}, {Conselice},
  {Crocce}, {Cunha}, {D'Andrea}, {da Costa}, {Davis}, {Desai}, {Diehl},
  {Dietrich}, {Dodelson}, {Doel}, {Drlica-Wagner}, {Estrada}, {Etherington},
  {Evrard}, {Fabbri}, {Finley}, {Flaugher}, {Foley}, {Fosalba}, {Frieman},
  {Garc{\'\i}a-Bellido}, {Gaztanaga}, {Gerdes}, {Giannantonio}, {Goldstein},
  {Gruen}, {Gruendl}, {Guarnieri}, {Gutierrez}, {Hartley}, {Honscheid}, {Jain},
  {James}, {Jeltema}, {Jouvel}, {Kessler}, {King}, {Kirk}, {Kron}, {Kuehn},
  {Kuropatkin}, {Lahav}, {Li}, {Lima}, {Lin}, {Maia}, {Makler}, {Manera},
  {Maraston}, {Marshall}, {Martini}, {McMahon}, {Melchior}, {Merson}, {Miller},
  {Miquel}, {Mohr}, {Morice-Atkinson}, {Naidoo}, {Neilsen}, {Nichol}, {Nord},
  {Ogando}, {Ostrovski}, {Palmese}, {Papadopoulos}, {Peiris}, {Peoples},
  {Percival}, {Plazas}, {Reed}, {Refregier}, {Romer}, {Roodman}, {Ross},
  {Rozo}, {Rykoff}, {Sadeh}, {Sako}, {S{\'a}nchez}, {Sanchez}, {Santiago},
  {Scarpine}, {Schubnell}, {Sevilla-Noarbe}, {Sheldon}, {Smith}, {Smith},
  {Soares-Santos}, {Sobreira}, {Soumagnac}, {Suchyta}, {Sullivan}, {Swanson},
  {Tarle}, {Thaler}, {Thomas}, {Thomas}, {Tucker}, {Vieira}, {Vikram},
  {Walker}, {Wechsler}, {Weller}, {Wester}, {Whiteway}, {Wilcox}, {Yanny},
  {Zhang}, \& {Zuntz}}]{DESsurvey}
{Dark Energy Survey Collaboration}, {Abbott}, T., {Abdalla}, F.~B., {et~al.}
  2016, \mnras, 460, 1270, \dodoi{10.1093/mnras/stw641}

\bibitem[{{Dawson} {et~al.}(2013){Dawson}, {Schlegel}, {Ahn}, {Anderson},
  {Aubourg}, {Bailey}, {Barkhouser}, {Bautista}, {Beifiori}, {Berlind},
  {Bhardwaj}, {Bizyaev}, {Blake}, {Blanton}, {Blomqvist}, {Bolton}, {Borde},
  {Bovy}, {Brandt}, {Brewington}, {Brinkmann}, {Brown}, {Brownstein}, {Bundy},
  {Busca}, {Carithers}, {Carnero}, {Carr}, {Chen}, {Comparat}, {Connolly},
  {Cope}, {Croft}, {Cuesta}, {da Costa}, {Davenport}, {Delubac}, {de Putter},
  {Dhital}, {Ealet}, {Ebelke}, {Eisenstein}, {Escoffier}, {Fan}, {Filiz Ak},
  {Finley}, {Font-Ribera}, {G{\'e}nova-Santos}, {Gunn}, {Guo}, {Haggard},
  {Hall}, {Hamilton}, {Harris}, {Harris}, {Ho}, {Hogg}, {Holder}, {Honscheid},
  {Huehnerhoff}, {Jordan}, {Jordan}, {Kauffmann}, {Kazin}, {Kirkby}, {Klaene},
  {Kneib}, {Le Goff}, {Lee}, {Long}, {Loomis}, {Lundgren}, {Lupton}, {Maia},
  {Makler}, {Malanushenko}, {Malanushenko}, {Mandelbaum}, {Manera}, {Maraston},
  {Margala}, {Masters}, {McBride}, {McDonald}, {McGreer}, {McMahon}, {Mena},
  {Miralda-Escud{\'e}}, {Montero-Dorta}, {Montesano}, {Muna}, {Myers},
  {Naugle}, {Nichol}, {Noterdaeme}, {Nuza}, {Olmstead}, {Oravetz}, {Oravetz},
  {Owen}, {Padmanabhan}, {Palanque-Delabrouille}, {Pan}, {Parejko},
  {P{\^a}ris}, {Percival}, {P{\'e}rez-Fournon}, {P{\'e}rez-R{\`a}fols},
  {Petitjean}, {Pfaffenberger}, {Pforr}, {Pieri}, {Prada}, {Price-Whelan},
  {Raddick}, {Rebolo}, {Rich}, {Richards}, {Rockosi}, {Roe}, {Ross}, {Ross},
  {Rossi}, {Rubi{\~n}o-Martin}, {Samushia}, {S{\'a}nchez}, {Sayres}, {Schmidt},
  {Schneider}, {Sc{\'o}ccola}, {Seo}, {Shelden}, {Sheldon}, {Shen}, {Shu},
  {Slosar}, {Smee}, {Snedden}, {Stauffer}, {Steele}, {Strauss}, {Streblyanska},
  {Suzuki}, {Swanson}, {Tal}, {Tanaka}, {Thomas}, {Tinker}, {Tojeiro},
  {Tremonti}, {Vargas Maga{\~n}a}, {Verde}, {Viel}, {Wake}, {Watson}, {Weaver},
  {Weinberg}, {Weiner}, {West}, {White}, {Wood-Vasey}, {Yeche}, {Zehavi},
  {Zhao}, \& {Zheng}}]{boss}
{Dawson}, K.~S., {Schlegel}, D.~J., {Ahn}, C.~P., {et~al.} 2013, \aj, 145, 10,
  \dodoi{10.1088/0004-6256/145/1/10}

\bibitem[{{de Jong} {et~al.}(2013){de Jong}, {Verdoes Kleijn}, {Kuijken}, \&
  {Valentijn}}]{KiDssurvey}
{de Jong}, J. T.~A., {Verdoes Kleijn}, G.~A., {Kuijken}, K.~H., \& {Valentijn},
  E.~A. 2013, Experimental Astronomy, 35, 25, \dodoi{10.1007/s10686-012-9306-1}

\bibitem[{{Dong} {et~al.}(2019){Dong}, {Zhang}, {Yu}, {Yang}, {Li}, {Han},
  {Luo}, {Zhang}, \& {Fu}}]{LDP}
{Dong}, F., {Zhang}, J., {Yu}, Y., {et~al.} 2019, \apj, 874, 7,
  \dodoi{10.3847/1538-4357/ab0648}

\bibitem[{{Fong} {et~al.}(2022){Fong}, {Han}, {Zhang}, {Yang}, {Gao}, {Wang},
  {Li}, {Katsianis}, \& {Alonso}}]{matt_ggl}
{Fong}, M., {Han}, J., {Zhang}, J., {et~al.} 2022, \mnras, 513, 4754,
  \dodoi{10.1093/mnras/stac1263}

\bibitem[{{Gaia Collaboration} {et~al.}(2018){Gaia Collaboration}, {Brown},
  {Vallenari}, {Prusti}, {de Bruijne}, {Babusiaux}, {Bailer-Jones}, {Biermann},
  {Evans}, {Eyer}, {Jansen}, {Jordi}, {Klioner}, {Lammers}, {Lindegren},
  {Luri}, {Mignard}, {Panem}, {Pourbaix}, {Randich}, {Sartoretti}, {Siddiqui},
  {Soubiran}, {van Leeuwen}, {Walton}, {Arenou}, {Bastian}, {Cropper},
  {Drimmel}, {Katz}, {Lattanzi}, {Bakker}, {Cacciari}, {Casta{\~n}eda},
  {Chaoul}, {Cheek}, {De Angeli}, {Fabricius}, {Guerra}, {Holl}, {Masana},
  {Messineo}, {Mowlavi}, {Nienartowicz}, {Panuzzo}, {Portell}, {Riello},
  {Seabroke}, {Tanga}, {Th{\'e}venin}, {Gracia-Abril}, {Comoretto},
  {Garcia-Reinaldos}, {Teyssier}, {Altmann}, {Andrae}, {Audard},
  {Bellas-Velidis}, {Benson}, {Berthier}, {Blomme}, {Burgess}, {Busso},
  {Carry}, {Cellino}, {Clementini}, {Clotet}, {Creevey}, {Davidson}, {De
  Ridder}, {Delchambre}, {Dell'Oro}, {Ducourant},
  {Fern{\'a}ndez-Hern{\'a}ndez}, {Fouesneau}, {Fr{\'e}mat}, {Galluccio},
  {Garc{\'\i}a-Torres}, {Gonz{\'a}lez-N{\'u}{\~n}ez}, {Gonz{\'a}lez-Vidal},
  {Gosset}, {Guy}, {Halbwachs}, {Hambly}, {Harrison}, {Hern{\'a}ndez},
  {Hestroffer}, {Hodgkin}, {Hutton}, {Jasniewicz}, {Jean-Antoine-Piccolo},
  {Jordan}, {Korn}, {Krone-Martins}, {Lanzafame}, {Lebzelter}, {L{\"o}ffler},
  {Manteiga}, {Marrese}, {Mart{\'\i}n-Fleitas}, {Moitinho}, {Mora}, {Muinonen},
  {Osinde}, {Pancino}, {Pauwels}, {Petit}, {Recio-Blanco}, {Richards},
  {Rimoldini}, {Robin}, {Sarro}, {Siopis}, {Smith}, {Sozzetti}, {S{\"u}veges},
  {Torra}, {van Reeven}, {Abbas}, {Abreu Aramburu}, {Accart}, {Aerts},
  {Altavilla}, {{\'A}lvarez}, {Alvarez}, {Alves}, {Anderson}, {Andrei},
  {Anglada Varela}, {Antiche}, {Antoja}, {Arcay}, {Astraatmadja}, {Bach},
  {Baker}, {Balaguer-N{\'u}{\~n}ez}, {Balm}, {Barache}, {Barata}, {Barbato},
  {Barblan}, {Barklem}, {Barrado}, {Barros}, {Barstow}, {Bartholom{\'e}
  Mu{\~n}oz}, {Bassilana}, {Becciani}, {Bellazzini}, {Berihuete}, {Bertone},
  {Bianchi}, {Bienaym{\'e}}, {Blanco-Cuaresma}, {Boch}, {Boeche}, {Bombrun},
  {Borrachero}, {Bossini}, {Bouquillon}, {Bourda}, {Bragaglia}, {Bramante},
  {Breddels}, {Bressan}, {Brouillet}, {Br{\"u}semeister}, {Brugaletta},
  {Bucciarelli}, {Burlacu}, {Busonero}, {Butkevich}, {Buzzi}, {Caffau},
  {Cancelliere}, {Cannizzaro}, {Cantat-Gaudin}, {Carballo}, {Carlucci},
  {Carrasco}, {Casamiquela}, {Castellani}, {Castro-Ginard}, {Charlot},
  {Chemin}, {Chiavassa}, {Cocozza}, {Costigan}, {Cowell}, {Crifo}, {Crosta},
  {Crowley}, {Cuypers}, {Dafonte}, {Damerdji}, {Dapergolas}, {David}, {David},
  {de Laverny}, {De Luise}, {De March}, {de Martino}, {de Souza}, {de Torres},
  {Debosscher}, {del Pozo}, {Delbo}, {Delgado}, {Delgado}, {Di Matteo},
  {Diakite}, {Diener}, {Distefano}, {Dolding}, {Drazinos}, {Dur{\'a}n},
  {Edvardsson}, {Enke}, {Eriksson}, {Esquej}, {Eynard Bontemps}, {Fabre},
  {Fabrizio}, {Faigler}, {Falc{\~a}o}, {Farr{\`a}s Casas}, {Federici},
  {Fedorets}, {Fernique}, {Figueras}, {Filippi}, {Findeisen}, {Fonti},
  {Fraile}, {Fraser}, {Fr{\'e}zouls}, {Gai}, {Galleti}, {Garabato},
  {Garc{\'\i}a-Sedano}, {Garofalo}, {Garralda}, {Gavel}, {Gavras}, {Gerssen},
  {Geyer}, {Giacobbe}, {Gilmore}, {Girona}, {Giuffrida}, {Glass}, {Gomes},
  {Granvik}, {Gueguen}, {Guerrier}, {Guiraud}, {Guti{\'e}rrez-S{\'a}nchez},
  {Haigron}, {Hatzidimitriou}, {Hauser}, {Haywood}, {Heiter}, {Helmi}, {Heu},
  {Hilger}, {Hobbs}, {Hofmann}, {Holland}, {Huckle}, {Hypki}, {Icardi},
  {Jan{\ss}en}, {Jevardat de Fombelle}, {Jonker}, {Juh{\'a}sz}, {Julbe},
  {Karampelas}, {Kewley}, {Klar}, {Kochoska}, {Kohley}, {Kolenberg},
  {Kontizas}, {Kontizas}, {Koposov}, {Kordopatis}, {Kostrzewa-Rutkowska},
  {Koubsky}, {Lambert}, {Lanza}, {Lasne}, {Lavigne}, {Le Fustec}, {Le
  Poncin-Lafitte}, {Lebreton}, {Leccia}, {Leclerc}, {Lecoeur-Taibi},
  {Lenhardt}, {Leroux}, {Liao}, {Licata}, {Lindstr{\o}m}, {Lister}, {Livanou},
  {Lobel}, {L{\'o}pez}, {Managau}, {Mann}, {Mantelet}, {Marchal}, {Marchant},
  {Marconi}, {Marinoni}, {Marschalk{\'o}}, {Marshall}, {Martino}, {Marton},
  {Mary}, {Massari}, {Matijevi{\v{c}}}, {Mazeh}, {McMillan}, {Messina},
  {Michalik}, {Millar}, {Molina}, {Molinaro}, {Moln{\'a}r}, {Montegriffo},
  {Mor}, {Morbidelli}, {Morel}, {Morris}, {Mulone}, {Muraveva}, {Musella},
  {Nelemans}, {Nicastro}, {Noval}, {O'Mullane}, {Ord{\'e}novic},
  {Ord{\'o}{\~n}ez-Blanco}, {Osborne}, {Pagani}, {Pagano}, {Pailler},
  {Palacin}, {Palaversa}, {Panahi}, {Pawlak}, {Piersimoni}, {Pineau}, {Plachy},
  {Plum}, {Poggio}, {Poujoulet}, {Pr{\v{s}}a}, {Pulone}, {Racero}, {Ragaini},
  {Rambaux}, {Ramos-Lerate}, {Regibo}, {Reyl{\'e}}, {Riclet}, {Ripepi}, {Riva},
  {Rivard}, {Rixon}, {Roegiers}, {Roelens}, {Romero-G{\'o}mez}, {Rowell},
  {Royer}, {Ruiz-Dern}, {Sadowski}, {Sagrist{\`a} Sell{\'e}s}, {Sahlmann},
  {Salgado}, {Salguero}, {Sanna}, {Santana-Ros}, {Sarasso}, {Savietto},
  {Schultheis}, {Sciacca}, {Segol}, {Segovia}, {S{\'e}gransan}, {Shih},
  {Siltala}, {Silva}, {Smart}, {Smith}, {Solano}, {Solitro}, {Sordo}, {Soria
  Nieto}, {Souchay}, {Spagna}, {Spoto}, {Stampa}, {Steele},
  {Steidelm{\"u}ller}, {Stephenson}, {Stoev}, {Suess}, {Surdej}, {Szabados},
  {Szegedi-Elek}, {Tapiador}, {Taris}, {Tauran}, {Taylor}, {Teixeira},
  {Terrett}, {Teyssandier}, {Thuillot}, {Titarenko}, {Torra Clotet}, {Turon},
  {Ulla}, {Utrilla}, {Uzzi}, {Vaillant}, {Valentini}, {Valette}, {van Elteren},
  {Van Hemelryck}, {van Leeuwen}, {Vaschetto}, {Vecchiato}, {Veljanoski},
  {Viala}, {Vicente}, {Vogt}, {von Essen}, {Voss}, {Votruba}, {Voutsinas},
  {Walmsley}, {Weiler}, {Wertz}, {Wevers}, {Wyrzykowski}, {Yoldas},
  {{\v{Z}}erjal}, {Ziaeepour}, {Zorec}, {Zschocke}, {Zucker}, {Zurbach}, \&
  {Zwitter}}]{gaiadr2}
{Gaia Collaboration}, {Brown}, A.~G.~A., {Vallenari}, A., {et~al.} 2018, \aap,
  616, A1, \dodoi{10.1051/0004-6361/201833051}

\bibitem[{{Gatti} {et~al.}(2021){Gatti}, {Sheldon}, {Amon}, {Becker}, {Troxel},
  {Choi}, {Doux}, {MacCrann}, {Navarro-Alsina}, {Harrison}, {Gruen},
  {Bernstein}, {Jarvis}, {Secco}, {Fert{\'e}}, {Shin}, {McCullough}, {Rollins},
  {Chen}, {Chang}, {Pandey}, {Tutusaus}, {Prat}, {Elvin-Poole}, {Sanchez},
  {Plazas}, {Roodman}, {Zuntz}, {Abbott}, {Aguena}, {Allam}, {Annis}, {Avila},
  {Bacon}, {Bertin}, {Bhargava}, {Brooks}, {Burke}, {Carnero Rosell}, {Carrasco
  Kind}, {Carretero}, {Castander}, {Conselice}, {Costanzi}, {Crocce}, {da
  Costa}, {Davis}, {De Vicente}, {Desai}, {Diehl}, {Dietrich}, {Doel},
  {Drlica-Wagner}, {Eckert}, {Everett}, {Ferrero}, {Frieman},
  {Garc{\'\i}a-Bellido}, {Gerdes}, {Giannantonio}, {Gruendl}, {Gschwend},
  {Gutierrez}, {Hartley}, {Hinton}, {Hollowood}, {Honscheid}, {Hoyle}, {Huff},
  {Huterer}, {Jain}, {James}, {Jeltema}, {Krause}, {Kron}, {Kuropatkin},
  {Lima}, {Maia}, {Marshall}, {Miquel}, {Morgan}, {Myles}, {Palmese},
  {Paz-Chinch{\'o}n}, {Rykoff}, {Samuroff}, {Sanchez}, {Scarpine}, {Schubnell},
  {Serrano}, {Sevilla-Noarbe}, {Smith}, {Soares-Santos}, {Suchyta}, {Swanson},
  {Tarle}, {Thomas}, {To}, {Tucker}, {Varga}, {Wechsler}, {Weller}, {Wester},
  \& {Wilkinson}}]{DESY3shear}
{Gatti}, M., {Sheldon}, E., {Amon}, A., {et~al.} 2021, \mnras, 504, 4312,
  \dodoi{10.1093/mnras/stab918}

\bibitem[{{Giblin} {et~al.}(2021){Giblin}, {Heymans}, {Asgari}, {Hildebrandt},
  {Hoekstra}, {Joachimi}, {Kannawadi}, {Kuijken}, {Lin}, {Miller},
  {Tr{\"o}ster}, {van den Busch}, {Wright}, {Bilicki}, {Blake}, {de Jong},
  {Dvornik}, {Erben}, {Getman}, {Napolitano}, {Schneider}, {Shan}, \&
  {Valentijn}}]{KiDs1000shear}
{Giblin}, B., {Heymans}, C., {Asgari}, M., {et~al.} 2021, \aap, 645, A105,
  \dodoi{10.1051/0004-6361/202038850}

\bibitem[{{Hirata} \& {Seljak}(2003)}]{regauss}
{Hirata}, C., \& {Seljak}, U. 2003, \mnras, 343, 459,
  \dodoi{10.1046/j.1365-8711.2003.06683.x}

\bibitem[{{Hoekstra} \& {Jain}(2008)}]{2008ARNPS..58...99H}
{Hoekstra}, H., \& {Jain}, B. 2008, Annual Review of Nuclear and Particle
  Science, 58, 99, \dodoi{10.1146/annurev.nucl.58.110707.171151}

\bibitem[{{Hsieh} \& {Yee}(2014)}]{demp}
{Hsieh}, B.~C., \& {Yee}, H.~K.~C. 2014, \apj, 792, 102,
  \dodoi{10.1088/0004-637X/792/2/102}

\bibitem[{{Jarvis} \& {Jain}(2008)}]{Jarvis}
{Jarvis}, M., \& {Jain}, B. 2008, \jcap, 2008, 003,
  \dodoi{10.1088/1475-7516/2008/01/003}

\bibitem[{{Kaiser}(2000)}]{2000ApJ...537..555K}
{Kaiser}, N. 2000, \apj, 537, 555, \dodoi{10.1086/309041}

\bibitem[{{Li} {et~al.}(2021){Li}, {Zhang}, {Liu}, {Luo}, {Zhang}, {Dong},
  {Shen}, \& {Wang}}]{snrf}
{Li}, H., {Zhang}, J., {Liu}, D., {et~al.} 2021, \apj, 908, 93,
  \dodoi{10.3847/1538-4357/abcda3}

\bibitem[{{Li} {et~al.}(2022){Li}, {Miyatake}, {Luo}, {More}, {Oguri},
  {Hamana}, {Mandelbaum}, {Shirasaki}, {Takada}, {Armstrong}, {Kannawadi},
  {Takita}, {Miyazaki}, {Nishizawa}, {Plazas Malagon}, {Strauss}, {Tanaka}, \&
  {Yoshida}}]{HSCY3shear}
{Li}, X., {Miyatake}, H., {Luo}, W., {et~al.} 2022, \pasj, 74, 421,
  \dodoi{10.1093/pasj/psac006}

\bibitem[{{Li} {et~al.}(2023){Li}, {Zhang}, {Sugiyama}, {Dalal}, {Rau},
  {Mandelbaum}, {Takada}, {More}, {Strauss}, {Miyatake}, {Shirasaki}, {Hamana},
  {Oguri}, {Luo}, {Nishizawa}, {Takahashi}, {Nicola}, {Osato}, {Kannawadi},
  {Sunayama}, {Armstrong}, {Komiyama}, {Lupton}, {Lust}, {Miyazaki},
  {Murayama}, {Nishimichi}, {Okura}, {Price}, {Tait}, {Tanaka}, \&
  {Wang}}]{HSCY3cosmos}
{Li}, X., {Zhang}, T., {Sugiyama}, S., {et~al.} 2023, arXiv e-prints,
  arXiv:2304.00702, \dodoi{10.48550/arXiv.2304.00702}

\bibitem[{{Liu} {et~al.}(2023){Liu}, {Meng}, {Er}, {Fan}, {Kilbinger}, {Li},
  {Li}, {Schrabback}, {Scognamiglio}, {Shan}, {Tao}, {Ting}, {Zhang}, {Cheng},
  {Farrens}, {Fu}, {Hildebrandt}, {Kang}, {Kneib}, {Liu}, {Mellier},
  {Nakajima}, {Schneider}, {Starck}, {Wei}, {Wright}, \& {Zhan}}]{LiuDZ_2023}
{Liu}, D.~Z., {Meng}, X.~M., {Er}, X.~Z., {et~al.} 2023, \aap, 669, A128,
  \dodoi{10.1051/0004-6361/202243978}

\bibitem[{{Mandelbaum} {et~al.}(2018){Mandelbaum}, {Miyatake}, {Hamana},
  {Oguri}, {Simet}, {Armstrong}, {Bosch}, {Murata}, {Lanusse}, {Leauthaud},
  {Coupon}, {More}, {Takada}, {Miyazaki}, {Speagle}, {Shirasaki}, {Sif{\'o}n},
  {Huang}, {Nishizawa}, {Medezinski}, {Okura}, {Okabe}, {Czakon}, {Takahashi},
  {Coulton}, {Hikage}, {Komiyama}, {Lupton}, {Strauss}, {Tanaka}, \&
  {Utsumi}}]{HSCY1shear}
{Mandelbaum}, R., {Miyatake}, H., {Hamana}, T., {et~al.} 2018, \pasj, 70, S25,
  \dodoi{10.1093/pasj/psx130}

\bibitem[{{Nishizawa} {et~al.}(2020){Nishizawa}, {Hsieh}, {Tanaka}, \&
  {Takata}}]{HSCphotoz}
{Nishizawa}, A.~J., {Hsieh}, B.-C., {Tanaka}, M., \& {Takata}, T. 2020, arXiv
  e-prints, arXiv:2003.01511.
\newblock \doarXiv{2003.01511}

\bibitem[{{Rowe} {et~al.}(2015){Rowe}, {Jarvis}, {Mandelbaum}, {Bernstein},
  {Bosch}, {Simet}, {Meyers}, {Kacprzak}, {Nakajima}, {Zuntz}, {Miyatake},
  {Dietrich}, {Armstrong}, {Melchior}, \& {Gill}}]{galsim}
{Rowe}, B.~T.~P., {Jarvis}, M., {Mandelbaum}, R., {et~al.} 2015, Astronomy and
  Computing, 10, 121, \dodoi{10.1016/j.ascom.2015.02.002}

\bibitem[{{Shupe} {et~al.}(2005){Shupe}, {Moshir}, {Li}, {Makovoz}, {Narron},
  \& {Hook}}]{wcssip}
{Shupe}, D.~L., {Moshir}, M., {Li}, J., {et~al.} 2005, in Astronomical Society
  of the Pacific Conference Series, Vol. 347, Astronomical Data Analysis
  Software and Systems XIV, ed. P.~{Shopbell}, M.~{Britton}, \& R.~{Ebert}, 491

\bibitem[{{van Uitert} {et~al.}(2017){van Uitert}, {Hoekstra}, {Joachimi},
  {Schneider}, {Bland-Hawthorn}, {Choi}, {Erben}, {Heymans}, {Hildebrandt},
  {Hopkins}, {Klaes}, {Kuijken}, {Nakajima}, {Napolitano}, {Schrabback},
  {Valentijn}, \& {Viola}}]{2017MNRAS.467.4131V}
{van Uitert}, E., {Hoekstra}, H., {Joachimi}, B., {et~al.} 2017, \mnras, 467,
  4131, \dodoi{10.1093/mnras/stx344}

\bibitem[{{Wang} {et~al.}(2023){Wang}, {Zhang}, {Li}, \& {Liu}}]{Haoran_pdf}
{Wang}, H., {Zhang}, J., {Li}, H., \& {Liu}, C. 2023, \apj, 954, 193,
  \dodoi{10.3847/1538-4357/acea76}

\bibitem[{{Wang} {et~al.}(2022){Wang}, {Yang}, {Zhang}, {Li}, {Fong}, {Xu},
  {He}, {Gu}, {Luo}, {Dong}, {Wang}, {Li}, {Katsianis}, {Wang}, {Shen}, {Alonso
  Vaquero}, {Liu}, {Huang}, \& {Liu}}]{Jiaqi_ggl}
{Wang}, J., {Yang}, X., {Zhang}, J., {et~al.} 2022, \apj, 936, 161,
  \dodoi{10.3847/1538-4357/ac8986}

\bibitem[{{Zhang} {et~al.}(2022){Zhang}, {Liu}, {Vaquero}, {Li}, {Wang},
  {Shen}, \& {Dong}}]{FQ_decals}
{Zhang}, J., {Liu}, C., {Vaquero}, P.~A., {et~al.} 2022, \aj, 164, 128,
  \dodoi{10.3847/1538-3881/ac84d8}

\bibitem[{{Zhang} {et~al.}(2015){Zhang}, {Luo}, \& {Foucaud}}]{FQ_shear}
{Zhang}, J., {Luo}, W., \& {Foucaud}, S. 2015, \jcap, 2015, 024,
  \dodoi{10.1088/1475-7516/2015/01/024}

\bibitem[{{Zhang} {et~al.}(2017){Zhang}, {Zhang}, \& {Luo}}]{FQ_pdf}
{Zhang}, J., {Zhang}, P., \& {Luo}, W. 2017, \apj, 834, 8,
  \dodoi{10.3847/1538-4357/834/1/8}

\bibitem[{{Zhang} {et~al.}(2019){Zhang}, {Dong}, {Li}, {Li}, {Li}, {Liu},
  {Luo}, {Fu}, {Li}, \& {Fan}}]{FQ_fd}
{Zhang}, J., {Dong}, F., {Li}, H., {et~al.} 2019, \apj, 875, 48,
  \dodoi{10.3847/1538-4357/ab1080}

\end{thebibliography}
\bibliographystyle{aasjournal}

\end{document}